\documentclass[a4paper,10.5pt]{article}

\usepackage[greek,english]{babel}
\usepackage[utf8x]{inputenc}

\usepackage{amsmath,amssymb,amsthm}

\usepackage{ amsthm, amscd, amsfonts, amssymb, graphicx,tikz, color, environ}

\usepackage{cite}
\usepackage{indentfirst}
\usepackage{graphicx}
\usepackage{calligra}

\usepackage[pdftex,bookmarks,colorlinks]{hyperref}
\hypersetup{colorlinks=false}
\usepackage{braket}
\usepackage{epsfig}

\usepackage{color}

\newcommand{\sgn}{\mathrm{sgn}}

\newcommand{\e}{\epsilon}

\newcommand{\M}{\textnormal{\tiny \textsc{m}}}
\newcommand{\W}{\textnormal{\tiny \textsc{w}}}
\newcommand{\ci}{\textnormal{\tiny \textsc{$\, \circ\, $}}}

\newcommand{\si}{\sigma}

\newcommand{\wh}{W^{\epsilon}}
\newcommand{\hb}{\epsilon}
\newcommand{\vh}{v^{\epsilon}}

\def\R{{\Bbb R}}

\def\e{\epsilon}

\def\L2{L^2}

\newcommand{\hf}{\widehat{f}}

\newtheorem{theorem}{Theorem}[section]

\theoremstyle{definition}

\theoremstyle{remark}
\newtheorem{remark}[theorem]{Remark}

\numberwithin{equation}{section}

\begin{document}

\title{An approximate series solution of the semiclassical Wigner equation}

\author{K.S. Giannopoulou \footnote{Department of Mathematical Sciences, Norwegian University of Science and Technology, Trondheim, Norway.
Email: {\bf konstantina.giannopoulou@ntnu.no}}
\footnote{{\bf Acknowledgement}
This work has been partially supported by the ``Maria Michail Manasaki" Bequest Fellowship \& the ERCIM ``Alain Bensoussan Fellowship Programme"
} \& G.N. Makrakis \footnote{Department of Mathematics \& Applied Mathematics,
University of Crete, Heraklion, Crete,
Greece \& Institute of Applied and Computational Mathematics, 
FORTH, Heraklion, Crete, Greece. Email: {\bf makrakg@iacm.forth.gr}} }







\maketitle
\begin{abstract}
We propose a new approximate series solution of the semiclassical Wigner equation by uniformization of WKB approximations
of the Schr\"odinger eigenfunctions.
\\
\\
{\bf  Keywords}  Schr\"odinger equation, Wigner equation, semiclassical limit, geometric optics, caustics, Weyl
quantization, Weyl symbols, uniform stationary phase method
\\
AMS (MOS) subject classification: 78A05, 81Q20, 53D55, 81S30, 34E05, 58K55
\end{abstract}
\tableofcontents

\section{Introduction}

We consider  the Cauchy problem for the 1-d Schr\"odinger equation with oscillatory  initial
data
\begin{equation}\label{1.1}
i\e u^{\epsilon}_t (x,t) =  -\frac{\e^2}{2}u^{\epsilon}_{xx}(x,t) + V(x) u^{\epsilon}(x,t)\, ,  \quad x\in R_{x} \quad ,\, \, \, t\in [0,T) \ ,
\end{equation}
\begin{equation}\label{1.2}
u^{\epsilon}(x,t=0) = u^{\epsilon}_{0}(x)= A_0 (x) \exp\bigl(i S_0 (x)/\e \bigr)\, ,
\end{equation}
where the potential $V(x) \in C^{\infty}(R_x)$ is real valued, $T$ is some positive constant,  $A_{0}(x)\in C_{0}^{\infty}(R_x)$, $S_{0}(x) \in C^{\infty}(R_x)$, and $\hb$ is a semiclassical parameter ($0<\hb<<1$), which plays the role of rescaled Planck's constant in quantum mechanics or the role of rescaled frequency in paraxial wave propagation.

\medskip
\noindent
{\bf Geometrical optics.}
When we are interested in the classical limit of quantum mechanics or in the propagation of high-frequency waves we have to study the limit of $u^{\epsilon}$ as $\e$ tends to zero.
This limit  has been traditionally studied by the method of geometrical optics (see, e.g., \cite{BLP}, \cite{BB}, \cite{KO1}). The method assumes the WKB ansatz
\begin{equation}\label{1.3}
u^{\epsilon}(x,t)= A(x,t) \exp\Bigl(iS(x,t)/\e \Bigr) \ ,  
\end{equation}
as the solution of the problem (\ref{1.1})-(\ref{1.2}).
Substituting (\ref{1.3}) into (\ref{1.1}), and retaining terms of order $O(1)$ in $\e$, we obtain that  the phase 
$S(x,t)$ satisfies the Hamilton-Jacobi  equation 
\begin{equation}\label{1.7}
S_t + H(x, \partial_x S) =0\, ,
\end{equation}
and the amplitude $A(x,t)$ satisfies the transport equation 
\begin{equation}\label{1.5}
2A_t + 2A_x S_x + AS_{xx} =0\, ,
\end{equation}
Here  $H(x, p)$ is the Hamiltonian function
\begin{equation}\label{1.6}
H(x, p) =p^{2}/2 + V(x) \ ,
\end{equation}
$p \in \R$ being the classical momentum.

Since the nonlinear equation (\ref{1.7}) does not in general have global in time solutions, the WKB  method  fails on caustics  where it predicts infinite wave amplitudes. From the mathematical point of view, formation of  caustics is associated to the multivaluedness of the phase function $S(x,t)$, and the crossing of rays. Formation of caustics  is a common situation in quantum mechanics and wave propagation  as a result of multipath propagation from localized sources. For example, even in the simplest oceanographic models and geophysical structures (see, e.g., \cite{TC},  \cite{CMP}),  various types of complicated caustics occur, depending upon the position of the sources and the stratification of the wave velocities. Several  techniques which exploit the geometry and properties of phase space for constructing  an ansatz in the form of a Fourier integral operator have been developed for studying oscillatory solutions  of wave equations and  the  homogenization of the corresponding energy densities  near caustics (see, e.g., \cite{MF}).

\medskip
\noindent
{\bf Wigner equation.} An entirely different approach  is, instead of assuming an ansatz, to introduce the Wigner transform (also reffered as Wigner function)
\begin{equation}\label{semWigner}
W^{\hb}[u^\hb](x,p,t)=(2\pi\hb)^{-1}\int_{R}^{} e^{-\frac{i}{\epsilon}y p} u^\hb\left(x+\frac{y}{2},t\right)\overline{u^\hb}\left(x-\frac{y}{2},t\right) \, dy 
\end{equation}
of the wave function $u^{\epsilon}$. 
Although the Wigner transform was introduced for the first time by Wigner \cite{Wig} in the context of quantum thermodynamics, as an alternative to missing quantum probability, it 
has been proved an extremely powerful tool for the construction of high-frequency asymptotics and the homogenization of energy densities of classical wave fields \cite{PR}.  The basic feature is that  the integration of the Wigner function against classical observables (symbols) with respect to the momentum $p$,  provides mean values of quantum observables or wave amplitude and energy flux of classical wave fields.

The Wigner transform has been proved a versatile tool in the 
reformulation of a wave equation as an integro-differential evolution equation (Wigner equation) which  governs the evolution of $W^{\hb}[u^\hb](x,p,t)$ in phase space. This approach has been proposed by Markowich and his collaborators \cite{Mark}, \cite{GMMP}, and it has been applied to several linear and non-linear evolution equations and systems. For example, assuming that the potential $V(x)$ in the Schr\"odinger equation $(\ref{1.1})$ is  smooth, the corresponding Wigner equation has the form
\begin{eqnarray}\label{wigner_ser}
&&\bigl({\partial_t} +p\partial_{x}-V'(x)\partial_{p}\bigr)W^\hb[u^\hb](x,p,t)\nonumber \\
\nonumber \\
&=&\sum_{n=1}^{\infty}\frac{(-1)^n}{2^{2n}(2n+1)!}\hb^{2n}
 V^{(2n+1)}(x) {\partial_p}^{2n+1}W^\hb[u^\hb](x,p,t) \ . 
\end{eqnarray}

In the high-frequency limit, the Wigner transform $(\ref{semWigner})$
converges weakly to the Wigner measure $W^{0}(x,p,t)$ \cite{LP}, which evolves according to the classical Liouville equation, 
$$\left({\partial_t} -V'(x)\partial_{p}+p\partial_{x}\right)W^0(x,p,t)=0 \ .$$
Liouville equation is the formal limit ($\hb=0$) of the Wigner equation $(\ref{wigner_ser})$, and it is equivalent to the WKB method only when no caustics appear.

However, it has been shown in \cite{FM1} that appropriate asymptotic expansion  of the Wigner transform \cite{Ber}, is able to produce caustic-free amplitudes in  certain cases of multipath propagation . This observation has motivated  the interest for the development of asymptotic solutions of the Wigner equation in the semiclassical regime $\hb<<1$. A perturbation solution of the Wigner equation has been constructed  in\cite{KalMak} for a class of single well potentials, by expanding the solution near  the Wigner function of the harmonic oscillator associated with the bottom of the well.

\medskip
\noindent
{\bf Scope of the paper.} In this paper we propose a new asymptotic approximation of the solution of the semiclassical Wigner equation, which is expressed as a series of Airy functions. 
We assume that the potential $V(x)$ is such that the spectrum of  the Schr\"odinger operator $\widehat{H}^{\epsilon}=-\frac{\epsilon ^2}{2}\frac{d^2}{dx^2}+V(x)$ be discrete. 
For simplicity we present here  the details of the construction in  the case  of the harmonic oscillator $V(x)={x^2}/{2}$. However, the same procedure can be applied for any potential well such that the spectrum of $\widehat{H}^{\epsilon}$ be discrete.

The construction of the approximate solution starts from the representation of the Wigner function by the eigenfunction series 
\begin{eqnarray}\label{wigner_exp}
W^{\hb}[u^\epsilon](x,p,t)=\sum_{n=0}^{\infty}\sum_{m=0}^{\infty} c_{nm}^{\hb}\, e^{-\frac{i}{\hb}(E_{n}^{\hb}-E_{m}^{\hb})t}\, W_{nm}^{\hb}(x,p) \ ,
\end{eqnarray}
where
$E_{n}^{\hb}$ are the eigenvalues of $\widehat{H}^{\epsilon}$,
and $\wh_{nm}$ are the {\it Wigner eigenfunctions}, that  is the Wigner transform of the eigenfunctions 
$v^\hb_{n}$ of $\widehat{H}^{\epsilon}$,
\begin{eqnarray}
\wh_{nm}(x,p)&=&W^\hb_{v_{m}^{\hb}}[v_{n}^{\hb}](x,p) \nonumber \\
&=&(2\pi\hb)^{-1}\int_{R}^{} e^{-\frac{i}{\epsilon}p y} v_{m}^{\hb}\left(x+\frac{y}{2}\right)\overline{v_{n}^{\hb}}\left(x-\frac{y}{2}\right)\, dy \ . \label{wigeigen}
\end{eqnarray}
The expansion coefficients are given by
\begin{eqnarray}\label{coeff_exp}
c_{nm}^{\hb}=
(W^\hb_{0},W_{nm}^{\hb})_{L^2({R}_{xp}^2)}
 \ , \ \ n,m=0,1,\ldots \ ,
\end{eqnarray}
where
\begin{eqnarray}\label{semWigner_ind}
W_{0}^{\hb}[u^\hb](x,p)&:=&W^{\hb}[u^{\hb}_{0}](x,p)\nonumber \\
&=&(\pi\hb)^{-1}\int_{R}^{} e^{-\frac{i}{2\epsilon} p\sigma} u^\hb_{0}\left(x+\sigma\right)\overline{u^\hb_{0}}\left(x-\sigma\right) \, d\sigma \ ,
\end{eqnarray}
is the Wigner transform of the initial wave function $u_{0}^{\hb}$.

The proposed  series  approximation of $W^{\hb}[u^\epsilon](x,p,t)$ is derived in three steps. 
First, $v^{\epsilon}_{n}$ in (\ref{wigeigen}) are replaced by their WKB approximations 
$\psi^{\epsilon}_{n}$, and $W_{nm}^{\hb}$ are approximated by the {\it WKB-Wigner eigenfunctions}
\begin{equation}\label{wigner_wkb_nm}
 \mathcal{W}^{\hb}_{nm}(x,p):=W^{\epsilon}_{\psi_m^\epsilon}[\psi_n^\epsilon](x,p)=(\pi\hb)^{-1}\int_{R}^{} e^{-\frac{i}{2\epsilon}p\sigma} \psi^{\epsilon}_{n}\left(x+\sigma\right)\overline{\psi^{\epsilon}_{m}}\left(x-\sigma\right) \, d\sigma \ ,
\end{equation}
Thus we substitute the expansion (\ref{wigner_exp}) by the approximation
\begin{eqnarray}\label{wigner_wkb_exp}
\mathcal{W}^{\hb}[u^\epsilon](x,p,t)=\sum_{n=0}^{\infty}\sum_{m=0}^{\infty} \mathcal{C}_{nm}^{\hb}\, e^{-\frac{i}{\hb}(E_{n}^{\hb}-E_{m}^{\hb})t}\, \mathcal{W}_{nm}^{\hb}(x,p) \ ,
\end{eqnarray}
where
\begin{eqnarray}\label{coeff_wkb_exp}
\mathcal{C}_{nm}^{\hb}=
(W^\hb_{0},\mathcal{W}_{nm}^{\hb})_{L^2({R}_{xp}^2)}
 \ , \ \ n,m=0,1,\ldots \ .
\end{eqnarray}
It can be shown by using WKB estimates for $(v^\hb_{n}-\psi^{\epsilon}_{n})$ (see, e.g., \cite{Fed}) that
\begin{eqnarray}\label{wf_exp_near}
\| W^{\hb}[u^\epsilon] - \mathcal{W}^{\hb}[u^\epsilon] \|_{L^{2}(R^{2}_{xp})} =o (\epsilon) \ ,  \ \ \mathrm{as} \ \ \epsilon \to 0 \ .
\end{eqnarray}
The exact form of the estimate $o (\epsilon)$ depends on the initial phase $S_{0}(x)$, through the dependence of $\widetilde{ \mathcal{C}_{nm}^{\hb}}$ on the initial phase as it will be shown by treating certain examples in Section 5.

Second,  the {\it uniformization procedure}, which was presented in \cite{GM} for a linear potential, is extended to the case of the harmonic oscillator, for the approximation of the WKB-Wigner eigenfunctions $\mathcal{W}^{\hb}_{nm}(x,p)$. For $m=n$ we derive the uniform semiclassical formula
\begin{eqnarray}\label{wigner_airapprox_nn}
{\mathcal{W}}^\epsilon_{n}(x,p) \approx && \widetilde{\mathcal{W}^{\hb}_{n}}(x,p):=
{\pi}^{-1}\, \hb^{-2/3}{\left(2E_{n}\right)}^{-1/3} \nonumber \\
&\times& Ai\left(\frac{p^2+x^2-2E_n}{\epsilon^{2/3}{(2E_{n})}^{1/3}}\right) \ .
\end{eqnarray}
For  large $n \ ,m \ , n \neq m$ and small $\hb$, such that $n\hb \ , m\hb$ and $(n-m)$ be fixed, we derive the uniform semiclassical formula
\begin{eqnarray}\label{wigner_airapprox_nm}
\mathcal{W}_{nm}^{\epsilon}(x,p)\approx \widetilde{\mathcal{W}^{\hb}_{nm}}(x,p):=
{\pi}^{-1}
{e^{-{i}(n-m)\phi(x,p)}}{\epsilon}^{-2/3}R_{nm}^{-4/3}(R_{nm}^2-\rho_{nm}^2)^{1/3} \nonumber  \\
\times Ai\left(\frac{p^2+x^2-R_{nm}^2}{\epsilon^{2/3}{R_{nm}^{4/3}(R_{nm}^2-\rho_{nm}^2)}^{-1/3}}\right) \ ,
\end{eqnarray}
where $Ai(.)$ is the Airy function, and
$$
\phi(x,p):=\arctan(p/x)\, , \quad R_{nm}=R^{\hb}_{nm}:=\frac{1}{2}(\sqrt{2E^{\hb}_{n}}+\sqrt{2E^{\hb}_{m}})$$
$$
\rho_{nm}=\rho^{\hb}_{nm}:=\frac{1}{2}(\sqrt{2E^{\hb}_{n}}-\sqrt{2E^{\hb}_{m}})\, .
$$
The uniformization procedure is based on the appropriate matching  (``asymptotic surgery") of various
local stationary-phase approximations of the integrals (\ref{wigner_wkb_nm}) in different regions of the phase space. 

Third, the expansion coefficients $\mathcal{C}_{nm}^{\hb}$ (eq. (\ref{coeff_wkb_exp})), are approximated by 
\begin{eqnarray}\label{approx_coeff_exp}
\widetilde{\mathcal{C}_{nm}^{\hb}}:=(\mathcal{W}^\hb_{0},\widetilde{\mathcal{W}_{nm}^{\hb}})_{L^2({R}_{xp}^2)}
\end{eqnarray}
where $\mathcal{W}^\hb_{0}$ is Berry's semiclassical Wigner function of the 
initial datum $W^\hb_{0}=W^{\hb}[u^\epsilon_{0}]$, which is given by (\ref{sclwigairy}) in Appendix C. 

Finally, the proposed approximation of $W^{\hb}[u^\epsilon](x,p,t)$ follows by substituting (\ref{approx_coeff_exp}) and (\ref {wigner_airapprox_nm}) into the eigenfunction series (\ref{wigner_wkb_exp}).

It is remarkable that the construction of the approximation  (\ref{wigner_airapprox_nm}) for the case of harmonic oscillator is  much more complicated than the construction of the corresponding approximation  for the linear potential in \cite{GM}. The main reason for this difference is the following.
In the case of the linear potential, the spectrum is continuous, and the Lagrangian curve in phase space is open (a parabola) with exactly one singular point, where the curve turns vertically. In the case of the harmonic oscillator, the spectrum is discrete, and we 
have an infinite  family of Lagrangian curves (circle), each having a couple of singular points. The interaction of these singular points makes the geometry of phase space very complicated and this affects tremendously the construction of the asymptotic formula.


\medskip
\noindent
{\bf Outline of the paper.}
The paper is organized as follows. In Section 1 we  present the method of geometrical optics, we illustrate how this method fails to predict the correct amplitudes on the caustics, and we introduced the semiclassical Wigner equation, which, according to recent literature is an alternative versatile tool in the study of high-frequency wave problems, avoiding caustic singularities. In Section 2 we introduce the Moyal star product, which is an efficient  tool for the derivation of  the Wigner equation from the Schr\"odinger equation, and for the construction of the eigenfunction series of the Wigner function in Section 3. The rest of the paper is concerned with the construction of an approximate solution of the semiclassical Wigner equation. The construction of the approximate solution of the semiclassical Wigner equation starts from an exact solution (eigenfunction series expansion) and it  employs Airy-type approximations of the Wigner eigenfunctions.
The eigenfunction series expansion is constructed in Section 3.2 using the spectral results of the Wigner operator in Section 3.1. In Section 4, which contains our main contribution, we describe with details
the proposed technique for the quadratic potential of the harmonic oscillator. Our approach is based on  Airy-type aapproximations of the Wigner eigenfunctions, which  are constructed by the uniformization of WKB functions. The most important details of the asymptotic constructions are presented  in Sections 4.2 and 4.3.
Finally, in Section 5 we present the approximate series of the Wigner function, we explain how the coefficients of the expansion can be calculated for certain initial data, and discuss some open questions concerning the calculation of the off-diagonal coefficients and the role of the incoherent part of the approximate series in the formation of caustics.

\section{Moyal star product and the Wigner equation} 

The construction of the eigenfunction series $(\ref{wigner_exp})$ relies on certain spectral results for the Wigner equation. These results will be deduced from some known results for the spectrum of an equation involving the so called Moyal star product, and for this reason we  need to write the Wigner equation in terms of this product.

The Moyal product is defined by the pseudo-differential operator 
\begin{equation}\label{moyal_product}
\star_{\M}:=\exp\Bigl[\frac{i\epsilon}{2}\Bigl(
\overleftarrow{\partial _x}
\overrightarrow{\partial _p}-
\overleftarrow{\partial _p}
\overrightarrow{\partial _x}\Bigr)\Bigr] \ .
\end{equation}
Considering the operators $\hf_{\W _1}^{\hb}$ and $\widehat{f}_{\W _2}^{\hb}$, with smooth  symbols  $f_1$ and$f_2$, respectively, in the Weyl representation, the symbol of the composition $\hf_{\W _1}^{\hb}\ci \widehat{f}_{\W _2}^{\hb}$, is given by (see, e.g., \cite{BS}, \cite{Zw}, \cite{Gro}, \cite{Tak}, \cite{Za1})
\begin{eqnarray*}\label{taylorprod}
(f_{1}\star_{\M} f_{2})(x,p)
&=&
\sum_{\alpha,\beta=1}^{\infty}\frac{(-1)^{\beta}}{\alpha !\beta!}{\left(\frac{i\epsilon}{2}\right)}^{\alpha+\beta}[\partial^{\alpha}_{x}\partial^{\beta}_{p}f_{1}(x,p)]
[\partial^{\beta}_{x}\partial^{\alpha}_{p}f_{2}(x,p)] \ .
\end{eqnarray*}

Let $\widehat{\rho}(t)$ be the pure-state density operator corresponding to the wave function $u^{\epsilon}$, which acts by 
$$\widehat{\rho}\psi= u^{\epsilon}(u^{\epsilon}, \psi) _{L^2({R}_{x}^2)} \ ,$$
for any $\psi \in L^2({R}_{x}^2)$. Then, the Schr\"odinger equation $(\ref{1.1})$ is equivalent to the Liouville-von Neumann equation (\cite{NSS}, \cite{KM})
\begin{eqnarray*}\label{vonneumann}
{i\hbar} \frac{d}{d t}\widehat{\rho}({t})=[\widehat{H}^{\hb},\widehat{\rho}(t)] \ ,
\end{eqnarray*}
where $[\widehat{H}^{\hb},\widehat{\rho}(t)]=\widehat{H}^{\hb}\ci\widehat{\rho}({t})-\widehat{\rho}({t})\ci\widehat{H}^{\hb}$ denotes the commutator of $\widehat{H}^{\hb}$ and $\widehat{\rho}(t)$. In the Weyl representation, the symbol of $\widehat{\rho}(t)$ is given by the Wigner transform $W^{\hb}[u^\epsilon](x,p,t)$, and the symbol of $\widehat{H}^{\hb}$ is the clasical Hamiltonian $H(x,p)$. Thus, equation $(\ref{vonneumann})$ implies the Wigner-Moyal equation
\begin{equation}\label{wignereq}
i\epsilon {\partial_t} W^\hb[u^\epsilon](x,p,t)=H(x,p)\star_{\M} W^\hb[u^\epsilon](x,p,t)-W^\hb[u^\epsilon](x,p,t)\star_{\M} H(x,p)
\end{equation}
for the symbols \cite{G1}.

By standard calculus for the Moyal product \cite{CFZ2}, we can rewrite equation $(\ref {wignereq})$ as the Wigner equation \footnote{In order to avoid notational confusion, we must mention that in mathematical literature (for example, in paraxial propagation of classical waves \cite{PR} and in homogenization theory \cite{GMMP}) it is customary to use the equation $(\ref{wignereqi})$, which is usually referred as the Wigner equation. On the other hand, physicists and mathematicians dealing with deformation quantization, prefer to work with the equation $(\ref {wignereq})$ which is referred as the Wigner-Moyal equation.}
\begin{eqnarray}\label{wignereqi}
{\partial_t} W^\hb[u^\hb](x,p,t)+\mathcal{L}^{\hb} W^\hb[u^\hb](x,p,t)=0 \ ,
\end{eqnarray}
where the quantum Liouville operator  $\mathcal{L}^{\hb}$ is defined in terms of the Hamiltonian  $H(x,p)$ and the sine operator (Moyal bracket \cite{Mo}, \cite{Gro}), 
\begin{eqnarray}\label{op_l}
\mathcal{L}^{\hb}\, \bullet:=-\frac{2}{\hb}\, H(x,p)\sin\left[\frac{\hb}{2}\left({\overleftarrow{\partial_
x}}\overrightarrow{\partial_p}-\overleftarrow{\partial_p}\overrightarrow{\partial_
x}\right)\right]\, \bullet \ .
\end{eqnarray}

By using the Taylor series of  sine into $\mathcal{L}^{\hb}$ we formally obtain
\begin{eqnarray*}
\sin\left[\frac{\hb}{2}\left({\overleftarrow{\partial_
x}}\overrightarrow{\partial_p}-\overleftarrow{\partial_p}\overrightarrow{\partial_
x}\right)\right]
=\frac{\hb}{2}\left({\overleftarrow{\partial_
x}}\overrightarrow{\partial_p}-\overleftarrow{\partial_p}\overrightarrow{\partial_
x}\right)+\sum_{n=1}^{\infty}\frac{(-1)^n}{(2n+1)!}\left(\frac{\hb}{2}\right)^{2n+1}\left({\overleftarrow{\partial_
x}}\overrightarrow{\partial_p}-\overleftarrow{\partial_p}\overrightarrow{\partial_
x}\right)^{2n+1} \ 
\end{eqnarray*}
and since the classical Hamiltonian $H(x,p)=\frac{p^2}{2}+V(x)$ is quadratic in $p$, equation$(\ref{wignereqi})$
reduces  to the Wigner equation $(\ref{wigner_ser})$, assuming that the potential $V(x)$ is smooth.

For  later use in the study of the spectrum of $\mathcal{L}^{\hb}$, it is also necessary to consider the cosine operator (Backer bracket \cite{Ba}, \cite{F})
\begin{eqnarray}\label{op_m}
\mathcal{M}^{\hb}\, \bullet:= H(x,p)\cos\left[\frac{\hb}{2}\left({\overleftarrow{\partial_
x}}\overrightarrow{\partial_p}-\overleftarrow{\partial_p}\overrightarrow{\partial_
x}\right)\right]\, \bullet \, .
\end{eqnarray}

It is important to note that the operators $\mathcal{L}^{\hb}$ and $\mathcal{M}^{\hb}$, arise naturally from the deformation quantization \cite{Bop}
$$x \rightarrow    x+\frac{i\epsilon}{2}\overrightarrow{\partial_ p}  \ , \ \ p \rightarrow p-\frac{i\epsilon}{2}\overrightarrow{\partial _x} \ ,$$
 of the classical Hamiltonian $H(x,p)$, to the deformed operator
\begin{eqnarray}\label{def_hamiltonian}
\mathbb{H}^\hb:=H\left(x+\frac{i\epsilon}{2}\overrightarrow{\partial_ p},p-\frac{i\epsilon}{2}\overrightarrow{\partial _x}\right) =\mathcal{M}^{\hb}-i\frac{\hb}{2}\mathcal{L}^{\hb}\ ,
\end{eqnarray}
which acts by 
\begin{eqnarray}\label{def_hamiltonian_action}
\mathbb{H}^\hb f(x,p) = H(x,p)\star_{\M}f(x,p) \ .
\end{eqnarray}
on any sufficiently smooth function $f$.
\section{Eigenfunction-series  solution of the Wigner equation}

In this section we construct an eigenfunction series expansion of the solution of the Cauchy problem
\begin{eqnarray}
i\epsilon {\partial_t} W^\hb[u^\epsilon](x,p,t)&=&H(x,p)\star_{\M} W^\hb[u^\epsilon](x,p,t)-W^\hb[u^\epsilon](x,p,t)\star_{\M} H(x,p) \label{wignereq_moy}\\
W^\hb_{0}[u^\epsilon](x,p)&=&W^\hb[u^\epsilon](x,p,0)=W^\hb[u^{\hb}_{0}](x,p)\label{wigner_ind}
\end{eqnarray}
where  
$W^\hb[u^{\hb}_{0}](x,p)$ is the semiclassical Wigner transform of initial wavefunction $u_{0}^{\epsilon}(x)$. This problem is the phase-space reformulation of (\ref{1.1})-(\ref{1.2}) in terms of the Wigner function.

Since the Wigner-Moyal  equation $(\ref{wignereq_moy})$ is linear, we apply the  method of separation of variables, and we look for a solution in the separated form 
\begin{equation}\label{wigner_sep}
W^\hb[u^\epsilon](x,p,t)=T^{\hb}(t)\Psi^{\hb}(x,p) \ . 
\end{equation}
Then, we get the equations 
\begin{eqnarray}
i\hb T^{\hb '}(t)={\mathcal{E}}^\hb T^{\hb}(t)\quad ,
\end{eqnarray}
and
\begin{eqnarray}\label{eq_spatial}
 H(x,p)\star_{\M} \Psi^{\epsilon}(x,p) -\Psi^{\epsilon}(x,p)\star_{\M} H(x,p)={\mathcal{E}}^{\hb}\Psi^{\hb}(x,p)
\end{eqnarray}
where ${\mathcal{E}}^{\hb}$ is the separation constant. The first equation has the solution   ${T^\hb}(t)=e^{-\frac{i}{\hb}{\mathcal{E}}^{\hb}t}$ up to a multiplicative constant. 
By using $(\ref{def_hamiltonian})$  and $(\ref{def_hamiltonian_action})$, the equation $(\ref{eq_spatial})$ is written in the form
 \begin{eqnarray*}
 (\mathbb{H}^\hb-\overline{\mathbb{H}^\hb})\Psi^{\hb}(x,p)=-i\hb\mathcal{L}^{\hb}\Psi_{}^{\hb}(x,p)={\mathcal{E}}^{\hb}\Psi^{\hb}(x,p) \ .
 \end{eqnarray*}
Therefore, we get the  eigenvalue problem 
\begin{eqnarray}\label{eigenl}
-i\hb\mathcal{L}^{\hb}\Psi_{}^{\hb}(x,p)={\mathcal{E}}^{\hb}\Psi_{}^{\hb}(x,p)\ ,
\end{eqnarray}
for the quantum Liouville operator $\mathcal{L}^{\hb}$.

We must note here that for solving the eigenvalue problem $(\ref{eigenl})$, it is necessary to solve simultaneously an associated eigenvalue problem for  $\mathcal{M}^{\hb}$ (see Section 3.2 below). It is also interesting to remark that the usual derivation of the semiclassical Wigner equation in paraxial propagation of classical and random waves \cite{PR}, where one starts from the Schr\"odinger equation in configuration space and uses the Wigner transform in an operational way, obscures the necessity and the role of the eigenvalue equation for  $\mathcal{M}^{\hb}$ since in this derivation 
$\mathcal{M}^{\hb}$ does not shop up at all. Therefore, the use of the formulation using the Moyal product seems indispensable.

\subsection{Eigenvalues of  $\mathcal{L}^{\hb}$ and $\mathcal{M}^{\hb}$}
The interrelation between the spectra of the
quantum Liouville operator $\mathcal{L}^{\hb}$  and the Schr\"odinger operator $\widehat{H}^\hb$
has been considered first by Spohn \cite{SP}, and it has
been further investigated and clarified by Antoniou et al. \cite{ASS}. They have shown, in a number of interesting cases, how the spectrum of $\mathcal{L}^{\hb}$ is  determined explicitly from that of $\widehat{H}^\hb$. Exploiting these results by Kalligiannaki \& Makrakis  \cite{KalMak}, have iddentified  the spectrum  $\mathcal{M}^{\hb}$.

In general, someone anticipates the formula 
$$\sigma(\mathcal{L}^\hb)=\left\{\frac{i}{\hb}(E^{\hb}-E^{\hb '})\, ,\quad E, \, E^{\hb '}\in \sigma(\widehat {H}^{\hb})\right\}
$$
to hold. In fact, this relation holds {\it for the discrete spectrum} 
\begin{equation}\label{eigl}
\si_{p}(\mathcal{L}^\hb)=\left\{\frac{i}{\hb}(E_{n}^{\hb}-E_{m}^{\hb})\, ,\quad E_{n}^{\hb},\,  E_{m}^{\hb}\in \sigma_p(\widehat {H}^{\hb})\right\} \ . 
\end{equation}
A similar formula holds for the point spectrum of the cosine braket operator $\mathcal{M}^\hb$ , that is 
\begin{equation}
\si_{p}(\mathcal{M}^\hb)=\left\{\frac{1}{2}(E_{n}^{\hb}+E_{m}^{\hb})\, ,\quad E_{n}^{\hb},\,  E_{m}^{\hb}\in \sigma_p(\widehat {H}^{\hb})\right\} \ . 
\end{equation}

However, these formulae are not in general true for the absolutely and singular continuous spectrum. These spectral questions have been studied first by Spohn \cite{SP} and later by Antoniou et al. \cite{ASS}, who have proved that
$$
\si_{sc,ac}(\mathcal {L}^{\hb})\neq \left\{ \frac{i}{\hb}(E^{\hb}-E^{\hb '})\, ,\quad E^{\hb},\,  E^{\hb '}\in \sigma_{sc,ac}(\widehat {H}^{\hb})\right\}\quad ,
$$
where $\si_{sc,ac}$ denote the singular and absolutely continuous spectrum.

In order to avoid the complications arising from the continuous spectrum (although this appears in the very interesting cases of scattering problems), we have considered operators $\widehat{H}^\hb$ with purely discrete spectrum $\si(\widehat{H}^{\hb})=\si_{p}(\widehat{H}^{\hb})$, in which case we have $\si(\mathcal{L}^{\hb})=\si_{p}(\mathcal{L}^{\hb})$ and  $\si(\mathcal{M}^{\hb})=\si_{p}(\mathcal{M}^{\hb})$ . 

\subsection{Eigenfunctions of  $\mathcal{L}^{\hb}$ and $\mathcal{M}^{\hb}$}
By straightforward computations, it has been shown by  Kalligiannaki \& Makrakis \cite{KalMak} that the Wigner eigenfunctions
$\{W_{nm}^{\hb}\}_{n,m=0,1,\ldots}$ (eq.$(\ref{wigeigen})$) form a complete orthonormal basis in $L^2(R_{xp}^{2})$ and they are common eigenfunctions of the operators  $\mathcal{L}^{\hb}$ and $\mathcal{M}^{\hb}$. 
\begin{theorem}

Let $\widehat{H}^{\hb}$ has purely discrete spectrum $\{E_{n}^\hb\}_{n=0,1,\dots}$  with complete orthonormal system of eigenfunctions $\{v_{n}^{\hb}(x)\}_{n=0,1,\dots}$ in $L^2(\R)$. Then, the functions $\{W_{nm}^{\hb}(x)\}_{n,m=0,1,\dots}$ (eq.$(\ref{wigeigen})$) form a complete orthonormal basis in $L^2(\R^{2}_{xp})$, and they are common eigenfunctions of operators  $\mathcal{L}^{\hb}$ and $\mathcal{M}^{\hb}$,
\begin{eqnarray}
\mathcal{L}^{\hb} W_{nm}^{\hb}(x,p)&=&\frac{i}{\hb}(E_{n}^{\hb}-E_{m}^{\hb})W_{nm}^{\hb}(x,p)\label{eig1} \label{eig1}\\
\mathcal{M}^{\hb} W_{nm}^{\hb}(x,p)&=&\frac{1}{2}(E_{n}^{\hb}+E_{m}^{\hb})W_{nm}^{\hb}(x,p)\label{eig2} \label{eig2}
\end{eqnarray}
in phase space $L^2(\R^{2}_{xp})$.
\end{theorem}

\begin{remark}
It has been shown by simple examples for linear and quadratic potential (\cite{CFZ1}, \cite{KP}, that for the computation of $\{W_{nm}^{\hb}\}$, someone must use both eigenvalue problems (\ref{eig1}) and (\ref{eig2}), and that only one of them is not enough for calculating the eigenfunction. On the other hand, it seems that there is not an evolution equation in phase space from which, by using separation of variables, someone can derive the eigenequation (\ref{eig2}). In an exciting attempt to accomplish this task, Fairlie \& Manogue  \cite{FMan} augmented the variables of the Wigner function by introducing an imaginary time  $s$. This leads to an evolution equation which involves the operator $i\partial_{s}+\mathcal{M}^{\hb}$ for the extended Wigner function, and it is formally quite similar to $(\ref{wignereqi})$. However, the study of this equation is still open.

\end{remark}

Comparing $(\ref{eigenl})$ with $(\ref{eig1})$
we see  the eigenvalue ${\mathcal{E}}^{\hb}$ takes the values $\mathcal{E}_{nm}^{\hb}:=E_{n}^{\hb}-E_{m}^{\hb}$,   and the corresponding eigenfunctions are given by $\Psi_{nm}^{\epsilon}(x,p)=\wh_{nm}(x,p)$. It is plausible to  argue that  $\mathcal{L}^{\hb}$ has not other eigenfunctions except  $\Psi_{nm}^{\epsilon}(x,p)=\wh_{nm}(x,p)$. To support this argument, by ${\mathcal{E}}^{\hb}=\mathcal{E}_{nm}^{\hb}$,  we write equation $(\ref{eq_spatial})$  in the form
\begin{eqnarray}\label{stat_eq_phs}
\left(H(x,p)\star_{\M} \Psi_{nm}^{\epsilon}(x,p) -E_{n}^{\hb}\Psi_{nm}^{\epsilon}(x,p)\right)-\left(\Psi_{nm}^{\epsilon}(x,p)\star_{\M} H(x,p)-E_{m}^{\hb}\Psi_{nm}^{\epsilon}(x,p)\right)=0 \ ,
\end{eqnarray}
and we make the plausible assumption that the  equations
\begin{eqnarray}\label{ass1}
H(x,p)\star_{\M} \Psi_{nm}^{\epsilon}(x,p) =E_{n}^{\hb}\Psi_{nm}^{\epsilon}(x,p)
\end{eqnarray}
\begin{eqnarray}\label{ass2}
\Psi_{nm}^{\epsilon}(x,p)\star_{\M} H(x,p)=E_{m}^{\hb}\Psi_{nm}^{\epsilon}(x,p)\ .
\end{eqnarray}
must hold simultaneously in order to $(\ref{stat_eq_phs})$  holds.

By Theorem 4 and Corollary 6 in \cite{GL}, it follows that all solutions of $(\ref{ass1})$ are given by $\Psi_{nm}^{\epsilon}=W_{nm}^{\epsilon}$. Then, we can check that  $\Psi_{nm}^{\epsilon}(x,p)=\wh_{nm}(x,p)$ satisfy the equation, $(\ref{ass2})$. Indeed,  by using the definition of the star exponential $(\ref{moyal_product})$, we have
$$\overline{W_{nm}^{\hb}\star_{\M}H}=\overline{W_{nm}^{\hb}\, e^{\frac{i\hb}{2}\left(
\overleftarrow{\partial _x}
\overrightarrow{\partial _p}-
\overleftarrow{\partial _p}
\overrightarrow{\partial _x}\right)}H}=H\,  e^{\frac{i\hb}{2}\left(
\overleftarrow{\partial _x}
\overrightarrow{\partial _p}-
\overleftarrow{\partial _p}
\overrightarrow{\partial _x}\right)}W_{mn}^{\hb}=H\star_{\M} W_{mn}^{\hb} \ .$$ 
Since the Wigner functions $W_{mn}^{\hb}(x,p)$ satisfy the eigenvalue equation  
$$H(x,p)\star_{\M}W_{mn}^{\hb}(x,p)=E_{m}W_{mn}^{\hb}(x,p) \ ,$$  
(see, e.g.,  \cite{CFZ1})
we have that $\overline{W_{nm}^{\hb}\star_{\M}H}=E_{m}W_{mn}^{\hb}(x,p)$, and  thus we get the equation
$(\ref{ass2})$.


\subsection{The eigenfunction series expansion}

By equations $(\ref{wigner_sep})$, $(\ref{eq_spatial})$, and the spectral results for $\mathcal{L}^{\hb}$ derived in the previous sections, the solution of Cauchy problem $(\ref{wignereq_moy})$-$(\ref{wigner_ind})$ is given by  eigenfunction  series (eq. $(\ref{wigner_exp})$)
\begin{eqnarray*}
W^{\hb}[u^\epsilon](x,p,t)=\sum_{n=0}^{\infty}\sum_{m=0}^{\infty} c_{nm}^{\hb}\, e^{-\frac{i}{\hb}(E_{n}^{\hb}-E_{m}^{\hb})t} W_{nm}^{\hb}(x,p)\ .
\end{eqnarray*}

The orthogonality of $W_{nm}^{\hb}$, implies that the coefficients $c_{nm}^{\hb}$ are given by the projections
(eq. $(\ref{coeff_exp})$) of the initial Wigner function $W_{0}^{\hb}[u^\hb]$  (eq. $(\ref{semWigner_ind})$) onto the Wigner eigenfunctions, 
\begin{eqnarray*}
c_{nm}^{\hb}=
(W_{0}^{\hb}[u^\hb],W_{nm}^{\hb})_{L^2(R_{xp}^2)} \  .
\end{eqnarray*}
Note that by the isometry of Wigner transform, the coefficients of the Wigner transform of the eigenfunction expansion can be expressed as
\begin{eqnarray*}
c_{nm}^{\hb}=(2\pi\hb)^{-1}(u_{0}^{\hb},v_{n}^{\hb})_{L^2(R_x)}\overline{(u_{0}^{\hb},v_{m}^{\hb})}_{L^2(R_x)}=
(W_{0}^{\hb}[u^\hb],W_{nm}^{\hb})_{L^2(R_{xp}^2)} \ .
\end{eqnarray*}

The eigenfunction series $(\ref{wigner_exp})$ can be written in the form 
\begin{equation}\label{wf_exp_cohdec}
W^{\hb}[u^\epsilon](x,p,t)= W^{\hb}[u^\epsilon]_{coh}(x,p) + W^{\hb}[u^\epsilon]_{incoh}(x,p,t)
\end{equation}
where
\begin{equation}\label{wf_exp_coh}
W^{\hb}[u^\epsilon]_{coh}(x,p):=\sum_{n=0}^{\infty}c_{nn}^{\hb}\,W_{nn}^{\hb}(x,p) \ ,
\end{equation}
and
\begin{equation}\label{wf_exp_incoh}
W^{\hb}[u^\epsilon]_{incoh}(x,p,t):=\sum _{n=0}^{\infty}\sum_{m=0, m\ne n}^{\infty} c_{nm}^{\hb}\, e^{-\frac{i}{\hb}(E_{n}^{\hb}-E_{m}^{\hb})t}\, W_{nm}^{\hb}(x,p) \ .
\end{equation}

The time-independent single series $(\ref{wf_exp_coh})$ is the {\it coherent part} of the solution of the Wigner equation, and it survives for all time. The time-dependent double series $(\ref{wf_exp_incoh})$ is the {\it  incoherent part} of the solution, which  is distributionally vanishing for large time, due to fast oscillations of the terms  $e^{-\frac{i}{\hb}(E_{n}^{\hb}-E_{m}^{\hb})t}$. This property  is referred as {\it decoherence}  in the theory of open quantum systems(see, e.g., \cite{BP}, Ch. 4). This means that  for large time, the solution of the Wigner equation, converges distributionally to the stationary solution $W^{\hb}[u^\epsilon]_{coh}$.
Moreover,  the incoherent part has zero net contribution to the wave energy, that is 
$$\int\int_{R^{2}_{xp} }W^{\hb}[u^\epsilon]_{incoh}(x,p,t)dxdp = 0 \ , $$
since 
$$\int\int_{R^{2}_{xp} }W_{nm}^{\hb}(x,p)dxdp=0 \ ,  \mathrm{for \ all } \
n \ , m = 0 \ 1 \ , \dots \ ,  \mathrm{with} \ n \ne m \ .$$
It turns out that the role of the off-diagonal Wigner eigenfunctions $W_{nm}^{\hb}$ is the exchange of energy between the modes of the solution.This mechanism seems to be crucial in understanding certain wave phenomena in the semiclassical regime and it will be further investigated elsewhere.

\begin{remark}
The expansion $(\ref{wigner_exp})$ has constructed for the first time in Moyal's pioneering paper \cite{Mo}, by applying the Wigner transform $(\ref{semWigner})$ term by term  onto the eigenfunction series expansion of the wave function $u^\epsilon$.

In the literature for deformation quantization  \cite{CFZ2},  it is customary to construct the solution of the Wigner-Moyal  equation $(\ref{wignereq_moy})$  by conjugating the initial data with the $\star$-unitary evolution operator,
$U_{\star_{\M}}(x,p;t)=e^{\frac{i}{\epsilon}tH}_{\star_{\M}}$, as follows 
\begin{eqnarray}\label{wignersolution}
W^\hb[u^\epsilon](x,p,t)=U_{\star_{\M}}^{-1}(x,p,t)\star_{\M} W^\hb_{0}[u^\epsilon](x,p)\star_{\M} U_{\star_{\M}}(x,p,t) \ .
\end{eqnarray}
The operator
$U_{\star_{\M}}(x,p,t)$  is represented through the star-resolution of identity by the series
\begin{eqnarray*}
U_{\star_{\M}}(x,p,t)=e^{\frac{it}{\epsilon}H}_{\star_{\M}}=e^{\frac{it}{\epsilon}H}_{\star_{\M}}\star_{\M} 1=
e^{\frac{it}{\epsilon}H}_{\star_{\M}}\star_{\M} 2\pi\hb\sum_{n=0}^{\infty}W_{nn}^{\hb}(x,p)=2\pi\hb\sum_{n=0}^{\infty}e^{itE_{n}/\hb}W_{nn}^{\hb}(x,p) \ ,
\end{eqnarray*}
which remarkably contains only the diagonal eigenfunctions $W_{nn}^{\hb}$ .

Then, by substituting  into $(\ref{wignersolution})$ the expansion of  $U_{\star_{\M}}(x,p,t)$, and of the initial datum $W^\hb_{0}[u^\epsilon](x,p)=\sum_{n=0}^{\infty}\sum_{m=0}^{\infty}c_{nm}^{\hb}W_{nm}^{\hb}(x,p) $, and using the projection formula 
\begin{eqnarray*}
&&(2\pi\hb) W_{mn}^{\hb}(x,p)\star_{\M} W_{k\ell}^{\hb}(x,p)=\delta_{nk} W_{m\ell}^{\hb}(x,p) \ ,
\end{eqnarray*}
someone leads to the expansion $(\ref{wigner_exp})$.

\end{remark}

\section{Airy approximation  of the Wigner eigenfunctions}

In this section we construct  Airy-type approximation of the Wigner eigenfunctions $W_{nm}^{\hb}$. These approximations are the necessary ingredients  for the construction an asymptotic series solution of the Wigner equation from  the eigenfunction series expansion $(\ref{wigner_exp})$. 

We present the details of the calculations for the simplest potential well, that is the  harmonic oscillator $V(x)=x^2/2$, for two reasons. First, in order to reduce the bulk of the asymptotic calculations and the phase-space geometrical complications. Second, because the Wigner eigenfunctions for the harmonic oscillator can be explicitly expressed in terms of special functions, with known asymptotics in certain regimes, and therefore we are able  to check the validity of the Airy approximations of the Wigner eigenfunctions. However, the same construction can be applied to any potential well which is non-degenerate at the bottom, and, in principle, it can be extended to higher dimensions, using canonical forms of the Hamiltonian functions \cite{CGsLRR}.

\subsection{The WKB-Wigner eigenfunctions}

Recall  from the Introduction that the starting point for the proposed construction is the approximation of the Wigner eigenfunctions $W_{nm}^{\hb}$ (eq.  (\ref{wigeigen})) by the \emph{WKB-Wigner eigenfunctions} $\mathcal{W}^{\hb}_{nm}$
\begin{eqnarray}\label{wigner_approx_nm_wkb}
\mathcal{W}^{\hb}_{nm}(x,p):=W^{\epsilon}_{\psi_m^\epsilon}[\psi_n^\epsilon]=(\pi\hb)^{-1}\int_{R}^{} e^{-\frac{i}{2\epsilon}p\sigma} \psi^{\epsilon}_{n}\left(x+\sigma\right)\overline{\psi^{\epsilon}_{m}}\left(x-\sigma\right) \, d\sigma 
\end{eqnarray}
for small $\hb$, where $\psi_n^\epsilon$ are the WKB approximations of eigenfunctions
$v^{\epsilon}_{n}$  of the Schr\"odinger operator $\widehat{H}^{\epsilon}=-\frac{\epsilon ^2}{2}\frac{d^2}{dx^2}+ \frac{x^2}{2}$. Using the two-phase representation of $\psi_n^\epsilon$ in the oscillatory region, see eqs. (\ref{wkbhosc}) , (\ref{wkbampl_1})  and  (\ref{wkbph_1}) in Appendix A, we rewrite 
 $(\ref{wigner_approx_nm_wkb})$ as a sum of four Wigner integrals, as follows.

\noindent
\subsection*{Diagonal WKB-Wigner  eigenfunctions $\mathcal{W}^{\hb}_{nn}$}
For $m=n$ we have

\begin{eqnarray} \label{eigenf_int}
{\mathcal{W}}^\epsilon_{n}(x,p)\equiv \mathcal{W}^{\hb}_{nn}(x,p)=
\sum_{\ell=1}^{4}\mathcal{W}^\epsilon_{\ell,n}(x,p)\nonumber\\
& &
\end{eqnarray}
where
\begin{equation}\label{wigner_int_n}
\mathcal {W}^\epsilon_{\ell,n}(x,p):=\frac{1}{\pi\epsilon}\int_{R}^{} D_{\ell,n}^{\hb}(\sigma;x)\,
e^{\frac{i}{\epsilon}F_{\ell,n}^{\hb}(\sigma;x,p)}\, d\sigma \quad , \quad \ell=1,\dots,4 \ .
\end{equation}
The amplitudes and phases of the above four Wigner integrals $\mathcal {W}^\epsilon_{\ell,n}$ are
given by
\begin{eqnarray}
D_{1,n}^{\hb}(\sigma;x)&=&{A_{n}^{\hb}}(x+\sigma){{A_{n}^\hb}}(x-\sigma) \ ,\nonumber \\
D_{2,n}^{\hb}(\sigma;x)&=&{A_{n}^\hb}(x+\sigma){{A_{n}^{\hb}}}(x-\sigma)  \ ,\nonumber \\
D_{3,n}^{\hb}(\sigma;x)&=&i{A_{n}^{\hb}}(x+\sigma){{A_{n}^\hb}}(x-\sigma)\ , \nonumber \\
D_{4,n}^{\hb}(\sigma;x)&=&-i{A_{n}^{\hb}}(x+\sigma){{A_{n}^\hb}}(x-\sigma) \ .
\end{eqnarray}
and 
\begin{eqnarray}
F_{1,n}^{\hb}(\sigma;x,p)&=&{S_{n}^\hb}(x+\sigma)-{S_{n}^\hb}(x-\sigma)-2p\sigma \ ,\label{phase_1n}\\
F_{2,n}^{\hb}(\sigma;x,p)&=&-\left({S_{n}^\hb}(x+\sigma)-{S_{n}^\hb}(x-\sigma)+2p\sigma \right) \ ,  \label{phase_2n}\\
F_{3,n}^{\hb}(\sigma;x,p)&=&{S_{n}^\hb}(x+\sigma)+{S_{n}^{\hb}}(x-\sigma)-2p\sigma \ , \label{phase_3n} \\
F_{4,n}^{\hb}(\sigma;x,p)&=&-\left({S_{n}^\hb}(x+\sigma)+{S_{n}^\hb}(x-\sigma)+2p\sigma \right)  \label{phase_4n} \ ,
\end{eqnarray}
where
\begin{eqnarray}
 A_{n}^{\hb}(x):=\frac{1}{2}\left(\frac{2}{\pi}\right)^{1/2}\left(2E_{n}^{\hb}-x^2 \right)^{-1/4}\ , \label{wkb_ampl} \\
  {S_{n}^{\hb}}(x):=\int_{\sqrt{2E_{n}^{\hb}}}^{x}\sqrt{2E_{n}^{\hb}-t^2}\,dt \ , \quad E_{n}^{\hb}=(n+1/2)\hb \ .\label{wkb_phase}
 \end{eqnarray}
We easily see that the Wigner phases $F_{\ell,n}^{\hb}$ satisfy the following relations 
\begin{eqnarray}
F_{2,n}^{\hb}(\sigma;x,p)&=&-F_{1,n}^{\hb}(\sigma;x,-p) \ , \label{diagsym_12} \\
F_{4,n}^{\hb}(\sigma;x,p)&=&-F_{3,n}^{\hb}(\sigma;x,-p) \ .   \label{diagsym_34}
\end{eqnarray}

\noindent
\subsection*{Off-diagonal WKB-Wigner  eigenfunctions $ \mathcal{W}^{\hb}_{nm}$}

For $m\ne n$ we have
\begin{equation}\label{wigner_int_nm}
{\mathcal{W}}_{nm}^\epsilon(x,p):=W^{\epsilon}_{\psi_m^\epsilon}[\psi_n^\epsilon](x,p)=
\sum_{\ell=1}^{4}\mathcal{W}^\epsilon_{\ell,nm}(x,p) \ ,
\end{equation}
where
\begin{equation}
\mathcal{W}^\epsilon_{\ell,nm}(x,p):=\int_{R}^{} D_{\ell,nm}^{\hb}(\sigma;x)\,
e^{\frac{i}{\epsilon}F_{\ell,nm}^{\hb}(\sigma;x,p)}\, d\sigma \, , \quad \ell=1,\ldots,4 \, , \quad n,m=0,1,\ldots \, .
\end{equation}
The amplitudes and phases of Wigner integrals $\mathcal {W}^\epsilon_{\ell,nm}$ are
given by

\begin{eqnarray}
D_{1,nm}^{\hb}(\sigma;x)&=&{A_{n}^\hb}(x+\sigma){{A_{m}^\hb}}(x-\sigma) \ , \nonumber \\
D_{2,nm}^{\hb}(\sigma;x)&=&{A_{n}^\hb}(x+\sigma){{A_{m}^\hb}}(x-\sigma)  \ , \nonumber \\
D_{3,nm}^{\hb}(\sigma;x)&=&i{A_{n}^\hb}(x+\sigma){{A_{m}^\hb}}(x-\sigma) \ , \nonumber \\
D_{4,nm}^{\hb}(\sigma;x)&=&-i{A_{n}^\hb}(x+\sigma){{A_{m}^\hb}}(x-\sigma)  \ .
\end{eqnarray}
and
\begin{eqnarray}
F_{1,nm}^{\hb}(\sigma;x,p)&=&{S_{n}^\hb}(x+\sigma)-{S_{m}^\hb}(x-\sigma)-2p\sigma \ , \label{phase_1nm}\\
F_{2,nm}^{\hb}(\sigma;x,p)&=&-\left({S_{n}^\hb}(x+\sigma)-{S_{m}^\hb}(x-\sigma)+2p\sigma \right) \ , \label{phase_2nm}\\
F_{3,nm}^{\hb}(\sigma;x,p)&=&{S_{n}^\hb}(x+\sigma)+{S_{m}^\hb}(x-\sigma)-2p\sigma \  , \label{phase_3nm} \\
F_{4,nm}^{\hb}(\sigma;x,p)&=&-\left({S_{n}^\hb}(x+\sigma)+{S_{m}^\hb}(x-\sigma)+2p\sigma \right) \ ,\label{phase_4nm}
\end{eqnarray}
where the amplitudes and phases, for $n$ and $m$, are given by  (\ref{wkb_ampl})-(\ref{wkb_phase}). Again, we see that the Wigner phases satisfy the relations 
\begin{eqnarray}
F_{2,nm}^{\hb}(\sigma;x,p)&=&-F_{1,nm}^{\hb}(\sigma;x,-p) \ , \label{offdiagsym_12} \\
F_{4,nm}^{\hb}(\sigma;x,p)&=&-F_{3,nm}^{\hb}(\sigma;x,-p) \ .  \label{offdiagsym_34}
\end{eqnarray}

\subsection{Airy approximation of $\mathcal{W}^{\epsilon}_{\ell,n}$}

Someone should keep in mind that the WKB approximation holds  for small $\epsilon$ and large enough $n$, so that the eigenvalue be considered constant, $E^{\hb}_{n}=(n+1/2)\hb\approx n\hb=:E_{n}$. Therefore, the Airy asymtptotics that we will construct in the sequel are subject to the same condition. For this reason, the amplitudes and phases
of the Wigner integrals are dimmed to be independent of $\hb$ and thus
we omit the superscript $\hb$ in the notations.

We first observe that the WKB phases ${S_{n}^{}}(x)$ are real in $|x|<\sqrt{2E_{n}^{}}$,   between the turning points, and become imaginary outside this set. Thus, the Wigner phases ${F}^{}_{\ell,n}$ , $\ell=1,2$ are imaginary in the set $\{\sigma\in \mathbb{R}: |x\pm\sigma|>\sqrt{2E_{n}^{}}\}$, and it is expected that the contribution of these stationary points to the integrals $(\ref{wigner_int_n})$ is exponentially small for small $\hb$.
Moreover, it turns out that all real stationary points of the Wigner phases lie in the set $\{\sigma\in \mathbb{R}: |x\pm\sigma|<\sqrt{2E_{n}^{}}\}$ where these phase are real. 

\subsubsection*{Construction of the asymptotics of $\mathcal{W}^{\epsilon}_{\ell,n}$ , $\ell=1,2$} 
\begin{figure}[h!]
\centering
\includegraphics[width=0.5\textwidth]{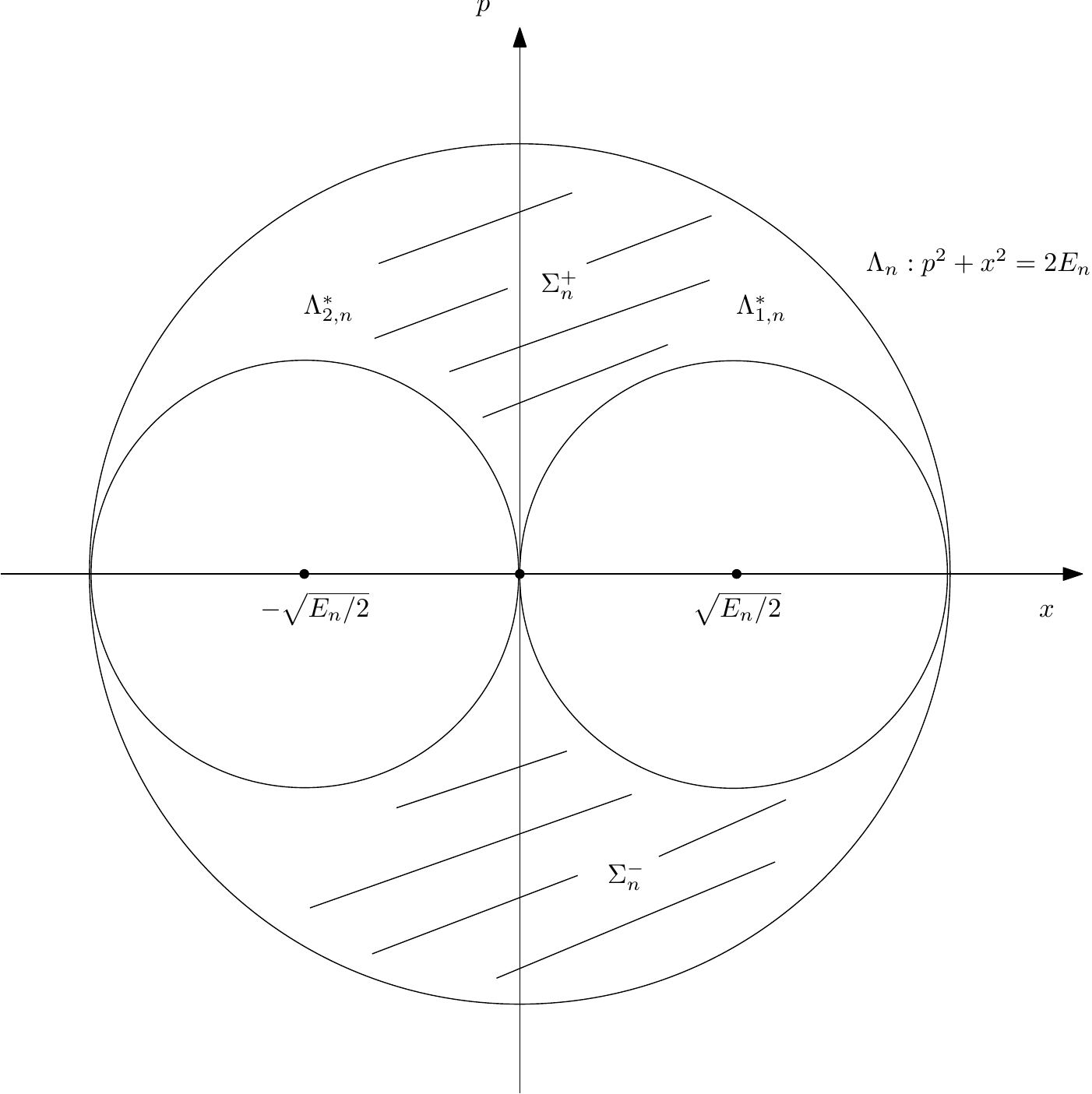}
\caption{{\it Lagrangian eigencurve $\Lambda_n$  and its dual curve $\Lambda^{*}_{n}$} }
\label{stat_points_n}
\end{figure}

The phases $F_{\ell,n}$ given by $(\ref{phase_1n})$-$(\ref{phase_2n})$, have a double stationary point $\si=0$, when $(x,p)$ lies on  the upper or on the lower branch of  Lagrangian eigencurve (see Fig. \ref{stat_points_n})
\begin{eqnarray}\label{curve_n}
\Lambda_{n}:=\lbrace (x,p)\in R^2: x^2 + p^2= 2E_{n}\rbrace \ .
\end{eqnarray}
defined by $H(x,p)= \frac12(x^2+p^2)=E_{n}$. 

Let
$$\Sigma_{n}=\Sigma_{n}^{+}\cup\Sigma_{n}^{-}$$
be the meniscus defined by the intersection of the interior of Lagrangian curve  $\Lambda_{n}$, and the exterior of its dual curve $\Lambda^{*}_{n}$. This curve is defined by
\begin{eqnarray}\label{union}
\Lambda^{*}_{n}:=\Lambda^{*}_{1,n}\cup \Lambda^{*}_{2,n} \ ,
\end{eqnarray}
where
\begin{eqnarray}\label{dual_n1}
\Lambda^{*}_{1,n}:=\lbrace (x,p)\in R^2:p^2+(x-\sqrt{E_{n}/2})^2= E_{n}/2\rbrace \ ,
\end{eqnarray}
and 
\begin{eqnarray}\label{dual_n2}
\Lambda^{*}_{2,n}:=\lbrace (x,p)\in R^2:p^2+(x+\sqrt{E_{n}/2})^2= E_{n}/2 \rbrace \ .
\end{eqnarray}

By the stationarity condition  $\partial_{\sigma}F_{\ell,n}=0$, we find that the phases $F_{\ell,n}$ ,  $\ell=1,2$, have a couple of real symmetric stationary points, given by
\begin{eqnarray}\label{sp}
\si(x,p)=\pm\frac{p}{\sqrt{p^2+x^2}}\sqrt{2E_{n}-p^2-x^2} \ , 
\end{eqnarray}
with $p>0$ when $(x,p)$ lies
in the upper meniscus $\Sigma_{n}^{+}$, and $p<0$ when $(x,p)$ lies in the lower meniscus $\Sigma_{n}^{-}$. There are not exist any stationary points in the interior of $\Lambda^{*}_{n}$.
Obviously the stationary points $(\ref{sp})$ coalesce to $\sigma(x,p)=0$ when  $(x,p)$  approaches $\Lambda_{n}$. Therefore, in this case we can use  Berry's semiclassical formula (\ref{sclwigairy}) in Appendix C.  By inspection of Berry's chord construction we identify the uniformity parameter $\alpha$ in (\ref{parameter_alpha}) by $\alpha=p-\sqrt{2E_{n}-x^2}$ when  $(x,p)$ lies in $\Sigma_{n}^{+}$, and $\alpha=p+\sqrt{2E_{n}-x^2}$ when  $(x,p)$ lies in $\Sigma_{n}^{-}$.

Then, by  (\ref{sclwigairy}) we get the approximation formula 
\begin{eqnarray}
\mathcal{W}_{1,n}^{\epsilon}(x,p)&\approx&
{\pi^{-1}}\, \hb^{-2/3}{\left(2E_{n}\right)}^{-1/3}
Ai\left(\frac{2\epsilon^{-2/3}}{\left(2E_{n}\right)^{1/3}\left(2E_{n}-x^2\right)^{-1/2}}\left(p-\sqrt{2E_{n}-x^2}\right)\right) \ ,\nonumber 
\end{eqnarray}
which,  is further approximated by 
\begin{eqnarray}\label{1n_appr}
\mathcal{W}_{1,n}^{\epsilon}(x,p)&\approx&{ \pi^{-1}}\, \hb^{-2/3} {\left(2E_{n}\right)}^{-1/3}\, Ai\left(\frac{ p^2+x^2-2E_n}{\hb^{2/3}{(2E_{n})}^{1/3}}\right) \ ,
\end{eqnarray}
near the upper branch $p=\sqrt{2E_{n}-x^2}$ of $\Lambda_{n}$, according to the uniformization procedure which we introduced in \cite{GM}.
Similarly we get the approximation
\begin{eqnarray}\label{2n_appr}
\mathcal{W}_{2,n}^{\epsilon}(x,p)\approx
{\pi^{-1}}\, \hb^{-2/3} {\left(2E_{n}\right)}^{-1/3}\, Ai\left(\frac{ p^2+x^2-2E_n}{\hb^{2/3}{(2E_{n})}^{1/3}}\right)  \ ,
\end{eqnarray}
near the lower branch $p=-\sqrt{2E_{n}-x^2}$ of  $\Lambda_{n}$. 

Away from  the branches $p=\pm\sqrt{2E_{n}-x^2}$ of $\Lambda_{n}$,  and inside 
$\Sigma_{n}^{\pm}$,  the approximations  $(\ref{1n_appr})$, $(\ref{2n_appr})$ are still valid, and, as it should be, they coincide with  the standard stationary-phase approximations of $\mathcal{W}^{\epsilon}_{\ell,n}$. In the interior of 
$\Lambda^{*}_{n}$, $\mathcal{W}^{\epsilon}_{\ell,n} \ , \ell=1 \ , 2$, are asymptotically small to any order of $\epsilon$.

We must note that the formulae $(\ref{1n_appr})$, $(\ref{2n_appr})$ are formally the same, but they are valid in different regions, since they have been derived by applying Berry's  semiclassical formula (\ref{sclwigairy}) near two different branches of the Lagrangian curve $\Lambda_{n}$.Thus,  for later use and notational convenience, we introduce the function
\begin{eqnarray}\label{new_notation}
\widetilde{\mathcal{W}^{\hb}_{n}}(x,p):= \pi^{-1}\, \hb^{-2/3}{\left(2E_{n}\right)}^{-1/3}\,
Ai\left(\frac{p^2+x^2-2E_n}{\epsilon^{2/3}{(2E_{n})}^{1/3}}\right) \  .
\end{eqnarray}

Outside $\Lambda_n$, where the stationary points of $F_{\ell,n}$, $\ell=1,2$ 
are imaginary (a pair) and again coalesce to a double
point on $\Lambda_n$ , we formally apply  the uniform stationary phase method (see eq. (\ref{ap4}) in Appendix B). Then, we get the approximation (\ref{new_notation}).

This procedure somehow corresponds to the analytic continuation of the asymptotic formulae in
the complex space. This is not rigorous since the Airy function becomes exponentially small
and the algebraic remainders cease to carry any approximation information.

\subsubsection*{Construction of the asymptotics of $\mathcal{W}^{\epsilon}_{\ell,n}$ , $\ell=3,4$}

The stationary points of the Wigner phases $F_{\ell,n}$ ,  $\ell=3,4$, are simple, when $(x,p)$ lies in the interior of $\Lambda^{*}_{n}$, and there not any  stationary points in $\Sigma_{n}^{\pm}$. Therefore, in this case we use the standard stationary phase formula (\ref{statphform}), and we derive the following approximations

\begin{itemize}
\item For $0\leqslant x<\sqrt{2E_{n}}$ and $p>0$ ,
\begin{eqnarray*}
\mathcal{W}_{3,n}^{\epsilon}(x,p)&\approx&\frac{1}{2\pi^{3/2}\sqrt{\epsilon}}\, 
e^{\frac{i}{\epsilon}F_{3,n}(-\sigma_{0})}e^{i\pi/4}(p^2+x^2)^{-1/4}(2E_{n}-p^2-x^2)^{-1/4}\quad ,\nonumber\\
\mathcal{W}_{4,n}^{\epsilon}(x,p)&\approx &\frac{1}{2\pi^{3/2}\sqrt{\epsilon}}\,
e^{-\frac{i}{\epsilon}F_{3,n}(-\sigma_{0})}e^{-i\pi/4}(p^2+x^2)^{-1/4}(2E_{n}-p^2-x^2)^{-1/4}\nonumber
\end{eqnarray*}
\item For $0\leqslant x<\sqrt{2E_{n}}$ and $p<0$ ,
\begin{eqnarray*}
\mathcal{W}_{3,n}^{\epsilon}(x,p)&\approx&\frac{1}{2\pi^{3/2}\sqrt{\epsilon}}\,
e^{\frac{i}{\epsilon}F_{3,n}(+\sigma_{0})}e^{i\pi/4}(p^2+x^2)^{-1/4}(2E_{n}-p^2-x^2)^{-1/4}\quad ,\nonumber\\
\mathcal{W}_{4,n}^{\epsilon}(x,p)&\approx&\frac{1}{2\pi^{3/2}\sqrt{\epsilon}}\,
e^{-\frac{i}{\epsilon}F_{3,n}(+\sigma_{0})}e^{-i\pi/4}(p^2+x^2)^{-1/4}(2E_{n}-p^2-x^2)^{-1/4}\nonumber
\end{eqnarray*}
\item For  $-\sqrt{2E_{n}}<x\leqslant 0$ and $p>0$ ,
\begin{eqnarray*}
\mathcal{W}_{3,n}^{\epsilon}(x,p)&\approx&\frac{1}{2\pi^{3/2}\sqrt{\epsilon}}\,
e^{\frac{i}{\epsilon}F_{3,n}(+\sigma_{0})}e^{i3\pi/4}(p^2+x^2)^{-1/4}(2E_{n}-p^2-x^2)^{-1/4}\quad , \nonumber\\
\mathcal{W}_{4,n}^{\epsilon}(x,p)&\approx&\frac{1}{2\pi^{3/2}\sqrt{\epsilon}}\,
e^{-\frac{i}{\epsilon}F_{3,n}(+\sigma_{0})}e^{-i3\pi/4}(p^2+x^2)^{-1/4}(2E_{n}-p^2-x^2)^{-1/4}\nonumber
\end{eqnarray*}
\item For $-\sqrt{2E_{n}}<x\leqslant 0$ and $p<0$ ,
\begin{eqnarray*}
\mathcal{W}_{3,n}^{\epsilon}(x,p)&\approx&\frac{1}{2\pi^{3/2}\sqrt{\epsilon}}\,
e^{\frac{i}{\epsilon}F_{3,n}(-\sigma_{0})}e^{i3\pi/4}(p^2+x^2)^{-1/4}(2E_{n}-p^2-x^2)^{-1/4} \ ,\nonumber\\
\mathcal{W}_{4,n}^{\epsilon}(x,p)&\approx&\frac{1}{2\pi^{3/2}\sqrt{\epsilon}}\,
e^{-\frac{i}{\epsilon}F_{3,n}(-\sigma_{0})}e^{-i3\pi/4}(p^2+x^2)^{-1/4}(2E_{n}-p^2-x^2)^{-1/4}  \nonumber
\end{eqnarray*}
\end{itemize}
where  $F_{3,n}(\pm\sigma_{0})=F_{3,n}(\si=\pm\sigma_{0}, x, p)$ with 
\begin{eqnarray*}
\si_{0}(x,p):=\frac{|p|}{\sqrt{p^2+x^2}}\sqrt{2E_{n}-p^2-x^2} \ .
\end{eqnarray*}
We observe that the approximation of $\mathcal{W}_{4,n}^{\epsilon}(x,p)$ is the complex conjugate of the approximation of $\mathcal{W}^{\epsilon}_{3,n}(x,p)$ . We must also note that in $\Sigma_{n}^{\pm}$, $\mathcal{W}^{\epsilon}_{\ell,n}(x,p) \ , \ell=3 \ , 4$,  are asymptotically small to any order of $\epsilon$.

By rather complicated transformations we can relate the phases $F_{3,n}(\pm \sigma_{0})$  to the phase of the Airy  approximation $(\ref {new_notation})$  of  
$\mathcal{W}_{\ell,n}^{\epsilon} \ , \ell=1,2$. These transformations have been motivated by  the asymptotics of Laguerre polynomials (see e.g. \cite{WoR}), which are the exact Wigner eigenfunctions for the harmonic oscillator. Then, by standard asymptotics of the Airy function, we can see that in the interior of  $\Lambda^{*}_{n}$ the sum  $(\mathcal{W}^{\hb}_{3,n}+\mathcal{W}^{\hb}_{4,n})$ matches with the asymptotics of $(\ref {new_notation})$, and therefore we get the approximation $(\mathcal{W}^{\hb}_{3,n}+\mathcal{W}^{\hb}_{4,n})\approx \widetilde{\mathcal{W}^{\hb}_{n}}$ .

Furthermore, by applying the so called complex stationary phase formula \cite{Tr}, (Ch.  X,  Sec. 3)  we show that 
\begin{eqnarray*}
\mathcal{W}_{3,n}^{\epsilon}(x,p)=0 \ , \ \ \ \mathcal{W}_{4,n}^{\epsilon}(x,p) = 0 \ 
\end{eqnarray*}
out of  $\Lambda_n$, since the imaginary stationary points contribute opposite terms in  the amplitudes due to phase jumps. These results  conform with the fact that, the approximation $(\ref{approx_w_n})$ is exponentially small in the exterior of $\Lambda_n$,  by the asymptotics of the Airy function.

\subsubsection*{The uniform approximation of $\mathcal{W}^{\hb}_{n}(x,p)$}

In Table \ref{table1} we summarize the contribution of different regions in the strip  $\{|x|<\sqrt{2E_{n}^{\hb}} \ , \  p \in \R\}$ to $\mathcal{W}^{\hb}_{n}$. The main contribution comes from the leading Airy term  $\widetilde{\mathcal{W}^{\hb}_{n}}$. 

\begin{center}
\begin{tabular}[!hbp]{|c|c|}\hline
region & main contribution to $\mathcal{W}^{\hb}_{n}(x,p)$ \\
\hline
$p>\sqrt{2E_{n}-x^2}$ & $\mathcal{W}^{\hb}_{1,n}\approx\widetilde{\mathcal{W}^{\hb}_{n}}$ \\
\hline
$p\approx\sqrt{2E_{n}-x^2}$ & $\mathcal{W}^{\hb}_{1,n}\approx\widetilde{\mathcal{W}^{\hb}_{n}}$ \\
\hline
inside $\Lambda^{*}_{n}$ & $\mathcal{W}^{\hb}_{3,n}+\mathcal{W}^{\hb}_{4,n} \approx\widetilde{\mathcal{W}^{\hb}_{n}}   $\\
\hline
$p\approx-\sqrt{2E_{n}-x^2}$ & $\mathcal{W}^{\hb}_{2,n}\approx\widetilde{\mathcal{W}^{\hb}_{n}}$ \\
\hline
$p>-\sqrt{2E_{n}-x^2}$ & $\mathcal{W}^{\hb}_{2,n}\approx\widetilde{\mathcal{W}^{\hb}_{n}}$ \\
\hline
\end{tabular}
\end{center}
\begin{table}[h]
\caption{\it{ The main contribution to ${\mathcal{W}}^\epsilon_{n}$} }
\label{table1}
\end{table}
Recall that   $\Lambda^{*}_{n}$ is the dual curve of $\Lambda_{n}$ (see eq. (\ref{union}) and Figure \ref{stat_points_n}), and  $\widetilde{\mathcal{W}^{\hb}_{n}}$ given by (\ref{new_notation}).
Thus, by the uniformization procedure \cite{GM}, the Airy approximation of  ${\mathcal{W}}^\epsilon_{n}$ is  given by
\begin{eqnarray}\label{approx_w_n}
{\mathcal{W}}^\epsilon_{n}(x,p) &\approx & \widetilde{\mathcal{W}^{\hb}_{n}}(x,p)\nonumber \\
&:=&
{\pi}^{-1}\, \hb^{-2/3}{\left(2E_{n}\right)}^{-1/3}\,
Ai\left(\frac{p^2+x^2-2E_n}{\epsilon^{2/3}{(2E_{n})}^{1/3}}\right)\quad .
\end{eqnarray}
for any $(x,p)$ in the strip.

\subsection{Airy approximation of  $\mathcal{W}^{\epsilon}_{\ell,n,m} $}

The construction of the asymptotics of  off-diagonal Wigner integrals $\mathcal{W}^\epsilon_{\ell,nm}$, $\ell=1,2$ (eq. (\ref{wigner_int_nm})), is much more complicated than that of the diagonal terms, because the double stationary points of $F_{\ell,nm}(\sigma;x,p)$   $\ell=1,2$, do not coalesce to zero on certain Lagrangian manifolds (contrary to what happens when $n=m$). For this reason we cannot use Berry's semiclassical Wigner function, but we must work with uniform stationary phase approximation (\ref{ap4}). 

Without loss of generality, suppose that $n>m$ and then $E_{n}^{\hb}>E_{m}^{\hb}$. In the construction of the approximation we assume that $n,m$ are large and $\epsilon$ small, so that $n\hb,m\hb=\mathrm{constant}$, and $n-m=\mathrm{constant}>0$. Then, the eigenvalues $E^{\hb}_{n}=(n+1/2)\hb\approx n\hb=:E_{n}$ and $E^{\hb}_{m}=(m+1/2)\hb\approx m\hb=:E_{m}$ can be treated as constants. 

\subsubsection*{Construction of the asymptotics of $\mathcal{W}^{\epsilon}_{\ell,n,m}$ , $\ell=1,2$}
\begin{figure}[h]
\centering
\includegraphics[width=0.5 \textwidth]{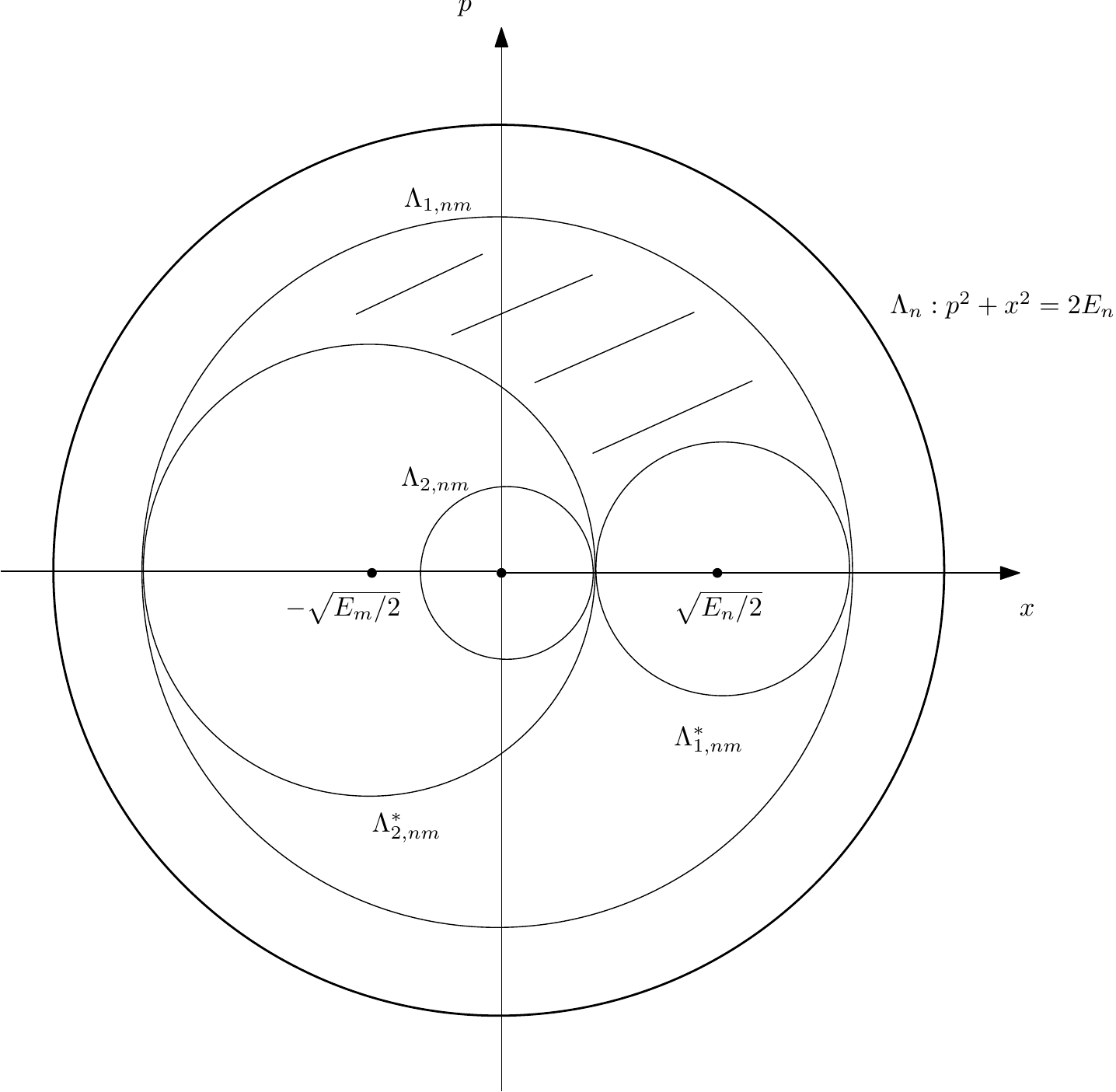}
\caption{{\it Area of existence of stationary points of $F_{1,nm} \   (\sigma=\si_{1,2} >0)$}}
\label{spnm_1a}
\end{figure}

\begin{figure}[h]
\centering
\includegraphics[width=0.5 \textwidth]{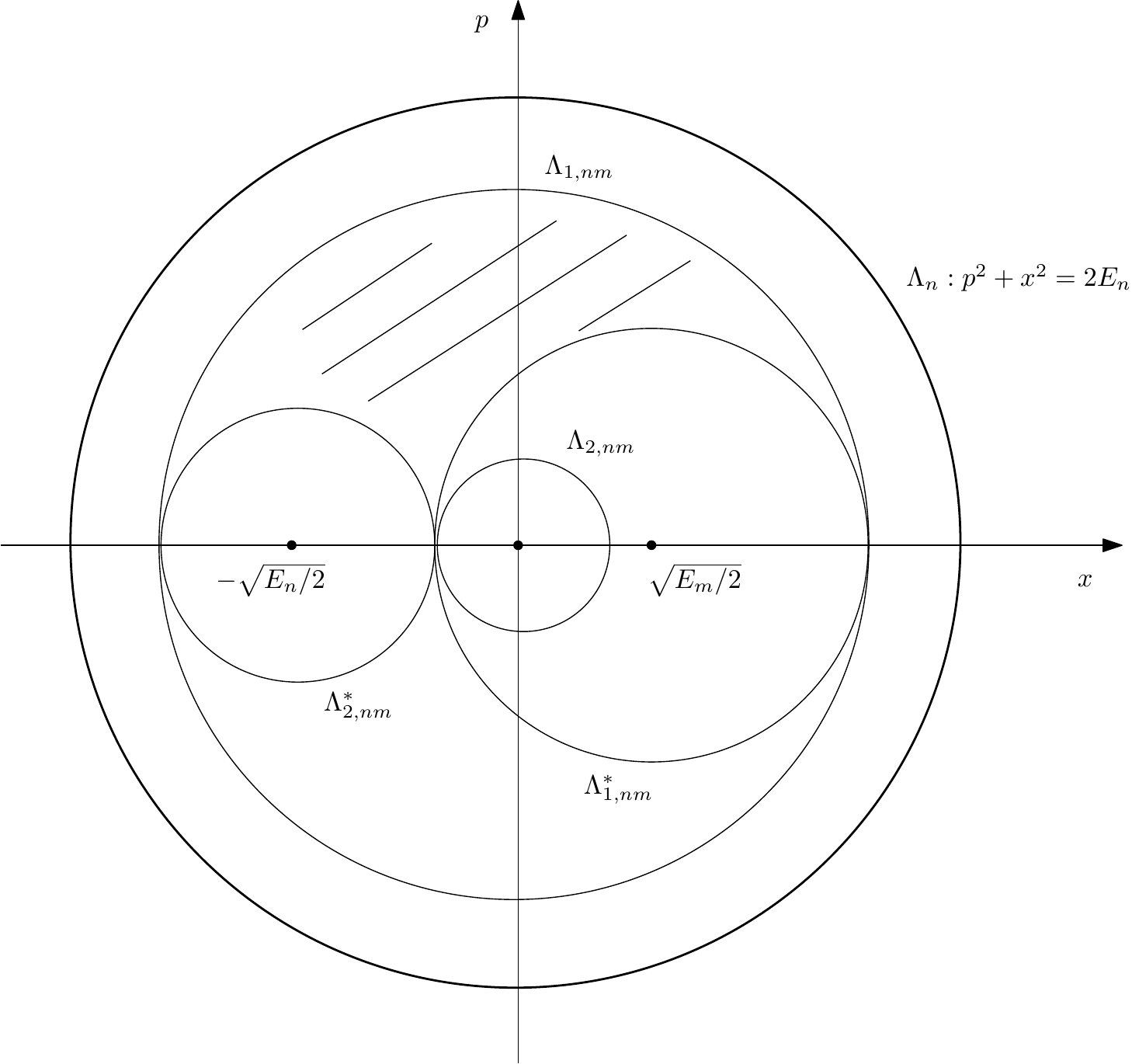}
\caption{{\it \it Area of existence of stationary points of $F_{1,nm} \  (\sigma=\si_{1,2} <0)$}}
\label{spnm_1b}
\end{figure}

By the stationarity condition $\partial_{\sigma}F_{\ell,nm}(\sigma,x,p)=0$ we find that
the stationary points are given by
\begin{eqnarray}\label{statpointsnm}
\si_{1,2}(x,p)=\frac{x e_{nm}}{p^2+x^2}\pm\frac{|p|}{p^2+x^2}\sqrt{(p^2+x^2)(2E_{nm}-p^2-x^2)-{e}^{2}_{nm}} \ ,
\end{eqnarray}
where
\begin{eqnarray}\label{Epsilon}
E_{nm}:=\frac{1}{2}\left(E_{n}+E_{m}\right)\quad \mathrm{and}\quad e_{nm}:=\frac{1}{2}\left(E_{n}-E_{m}\right) \ .
\end{eqnarray}
It turns out that these points are real when $(x,p)$ lies in the meniscus 
$$\Sigma_{nm}=\Sigma_{nm}^{+}\cup\Sigma_{nm}^{-} \ . $$
This meniscus is defined as the intersection of the ring $\rho^{2}_{nm} \le x^2+p^2 \leq R^{2}_{nm}$ lying between the Lagrangian curves 
 \begin{eqnarray}
\Lambda _{1,nm}&=&\lbrace (x,p)\in R^2:p^2+x^2=R_{nm}^2\rbrace\quad ,\label{man_nm_a} \\
\Lambda _{2,nm}&=&\lbrace (x,p)\in R^2:p^2+x^2=\rho_{nm}^2\rbrace\quad ,\label{man_nm_b}
\end{eqnarray}
where
\begin{eqnarray}
R_{nm}&:=&\frac{1}{2}(\sqrt{2E_{n}}+\sqrt{2E_{m}})\quad ,\label{z1}\\
\rho_{nm}&:=&\frac{1}{2}(\sqrt{2E_{n}}-\sqrt{2E_{m}}) \label{z2} 
\end{eqnarray}
and the exterior of their dual curve $\Lambda^{*}_{nm}$,
\begin{eqnarray}\label{union_nm}
\Lambda^{*}_{nm}:=\Lambda^{*}_{1,nm}\cup \Lambda^{*}_{2,nm} \ ,
\end{eqnarray}
where
\begin{itemize}
\item{ if $\sigma=\si_{1,2} \geq 0$ (see Figure \ref{spnm_1a})
\begin{eqnarray}\label{dual_n2}
\Lambda^{*}_{1,nm}:=\lbrace (x,p)\in R^2:   p^2+{(x-\sqrt{E_n/2})}^2 =E_{m}/2      \rbrace \ .
\end{eqnarray}
}
and
\begin{eqnarray}\label{dual_n1}
\Lambda^{*}_{2,nm}:=\lbrace (x,p)\in R^2:    p^2+{(x+\sqrt{E_m/2})}^2 = E_{n}/2   \rbrace \ ,
\end{eqnarray}
\item{if $\sigma=\si_{1,2} \leq 0$ (see Figure \ref{spnm_1b})
\begin{eqnarray}\label{dual_n1}
\Lambda^{*}_{1,nm}:=\lbrace (x,p)\in R^2:    p^2+{(x-\sqrt{E_m/2})}^2 = E_{n}/2   \rbrace \ ,
\end{eqnarray}
and 
\begin{eqnarray}\label{dual_n2}
\Lambda^{*}_{2,nm}:=\lbrace (x,p)\in R^2:   p^2+{(x+\sqrt{E_n/2})}^2 = E_{m}/2   \rbrace \ .
\end{eqnarray}
}
\end{itemize}
\newpage
Also, it turns out that there are no real stationary point for $(x,p)$ lying in the interior of $\Lambda _{2,nm}$ . 

The stationary points $(\ref{statpointsnm})$ coalesce to the double point
\begin{eqnarray}\label{si0bar}
\bar{\si}_0(x,p):=\frac{xe_{nm}}{p^2+x^2} \ ,
\end{eqnarray}
when $(x,p) \in \Lambda _{1,nm}\cup \Lambda _{2,nm}$. Thus, the curves $\Lambda _{1,nm} \ , \Lambda _{2,nm}$ may be considered as  the analogue of $\Lambda_n$ (eq. $(\ref{curve_n})$),which arouse in the diagonal case, since, formally, $\rho_{nm}=0$ and $R_{nm}=R_{n}:=2E_{n}$ for $n=m$.

Then, by applying the uniform stationary phase formula (\ref{ap4}) to $\mathcal{W}_{1,nm}^{\epsilon}$, we derive  the following approximation, for small $\hb$ and  $(x,p)$ near the branch $p=\sqrt{R^{2}_{nm}-x^2}>0$,
\begin{eqnarray*}
\mathcal{W}_{1,nm}^{\epsilon}(x,p)\approx
{\pi}^{-1}e^{-i(n-m)\phi}{\epsilon}^{-2/3}R_{nm}^{-4/3}(R_{nm}^2-\rho_{nm}^2)^{1/3}
Ai\left[\frac{p^2+x^2-R_{nm}^2}{\epsilon^{2/3}{R_{nm}^{4/3}(R_{nm}^2-\rho_{nm}^2)}^{-1/3}}\right]\, ,
\end{eqnarray*}
 where the angle $\phi$ is defined by
$\phi \equiv \phi(x,p):=\arctan(p/x)$, and $R_{nm}$, $r_{nm}$ given by $(\ref{z1})$ and $(\ref{z2})$, respectively.

In a similar way, for the second integral $\mathcal{W}^{\epsilon}_{2,nm}$, we derive the approximation
\begin{eqnarray*}
\mathcal{W}_{2,nm}^{\epsilon}(x,p)\approx {\pi}^{-1}
e^{-i(n-m)\phi}{\epsilon}^{-2/3}R_{nm}^{-4/3}(R_{nm}^2-\rho_{nm}^2)^{1/3}
Ai\left[\frac{p^2+x^2-R_{nm}^2}{\epsilon^{2/3}{R_{nm}^{4/3}(R_{nm}^2-\rho_{nm}^2)}^{-1/3}}\right]\, ,
\end{eqnarray*}
for small $\hb$ and near the branch $p=-\sqrt{R^{2}_{nm}-x^2}<0$ .
 
It must be emphasized that although the above two formulas are formally the same, they are valid for different values of $p$. Thus we are lead to define
\begin{eqnarray}\label{new_notation_nm}
\widetilde{\mathcal{W}_{nm}^{\epsilon}}(x,p):= {\pi}^{-1}
e^{-i(n-m)\phi}{\epsilon}^{-2/3}R_{nm}^{-4/3}(R_{nm}^2-\rho_{nm}^2)^{1/3}
Ai\left[\frac{p^2+x^2-R_{nm}^2}{\epsilon^{2/3}{R_{nm}^{4/3}(R_{nm}^2-\rho_{nm}^2)}^{-1/3}}\right] \ . \nonumber\\
\end{eqnarray}

\subsubsection*{Construction of the asymptotics of $\mathcal{W}^{\epsilon}_{\ell,n,m} \ , \ell=3 \ , 4$}

\begin{figure}[h]
\centering
\includegraphics[width=0.5 \textwidth]{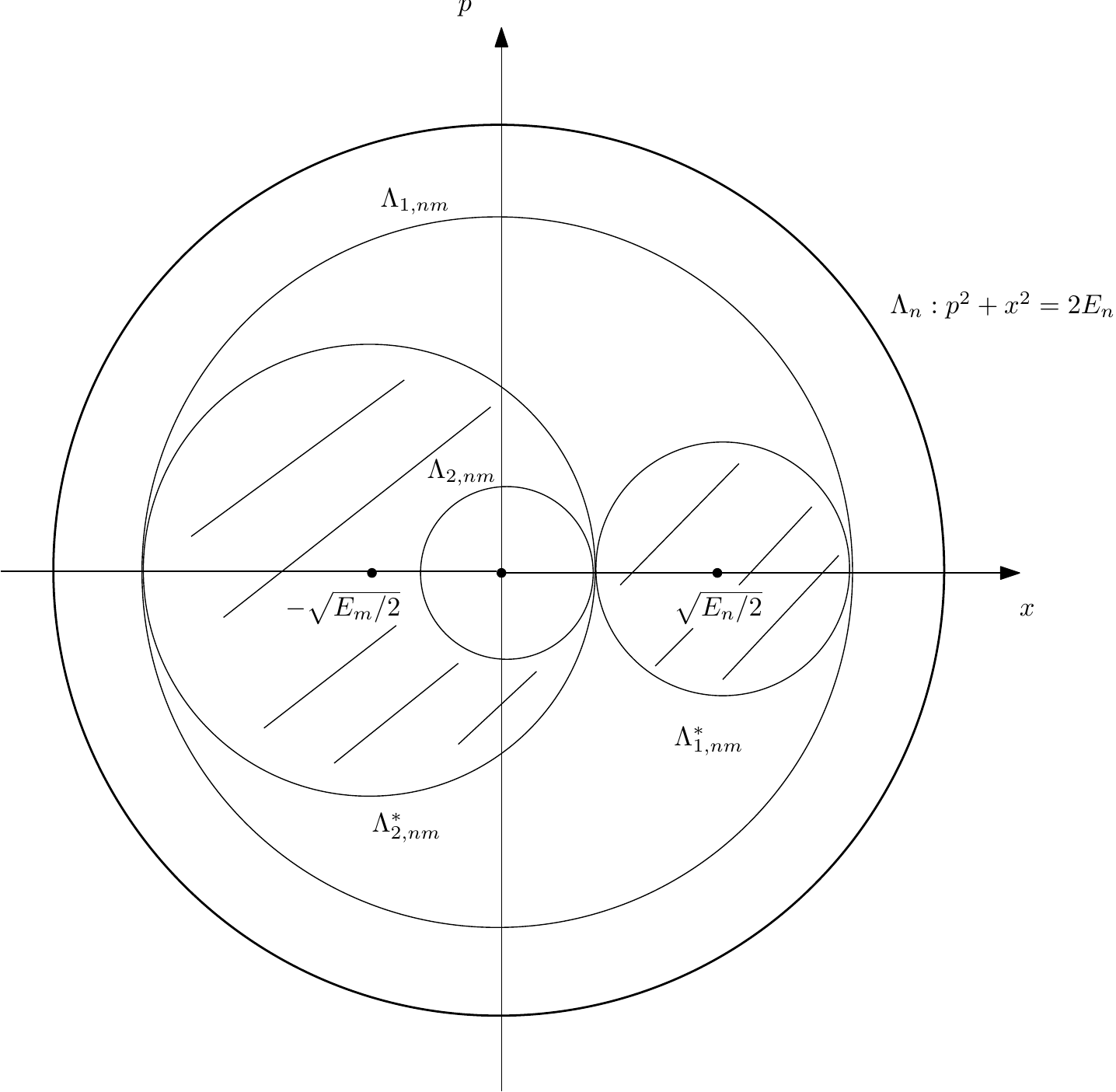}
\caption{{\it Area of existence of stationary points of $F_{\ell,nm}$, $\ell=3,4$  $(\sigma=\si_{1,2} >0)$}}
\label{stationaryp_34nma}
\end{figure}

\begin{figure}[h]
\centering
\includegraphics[width=0.7 \textwidth]{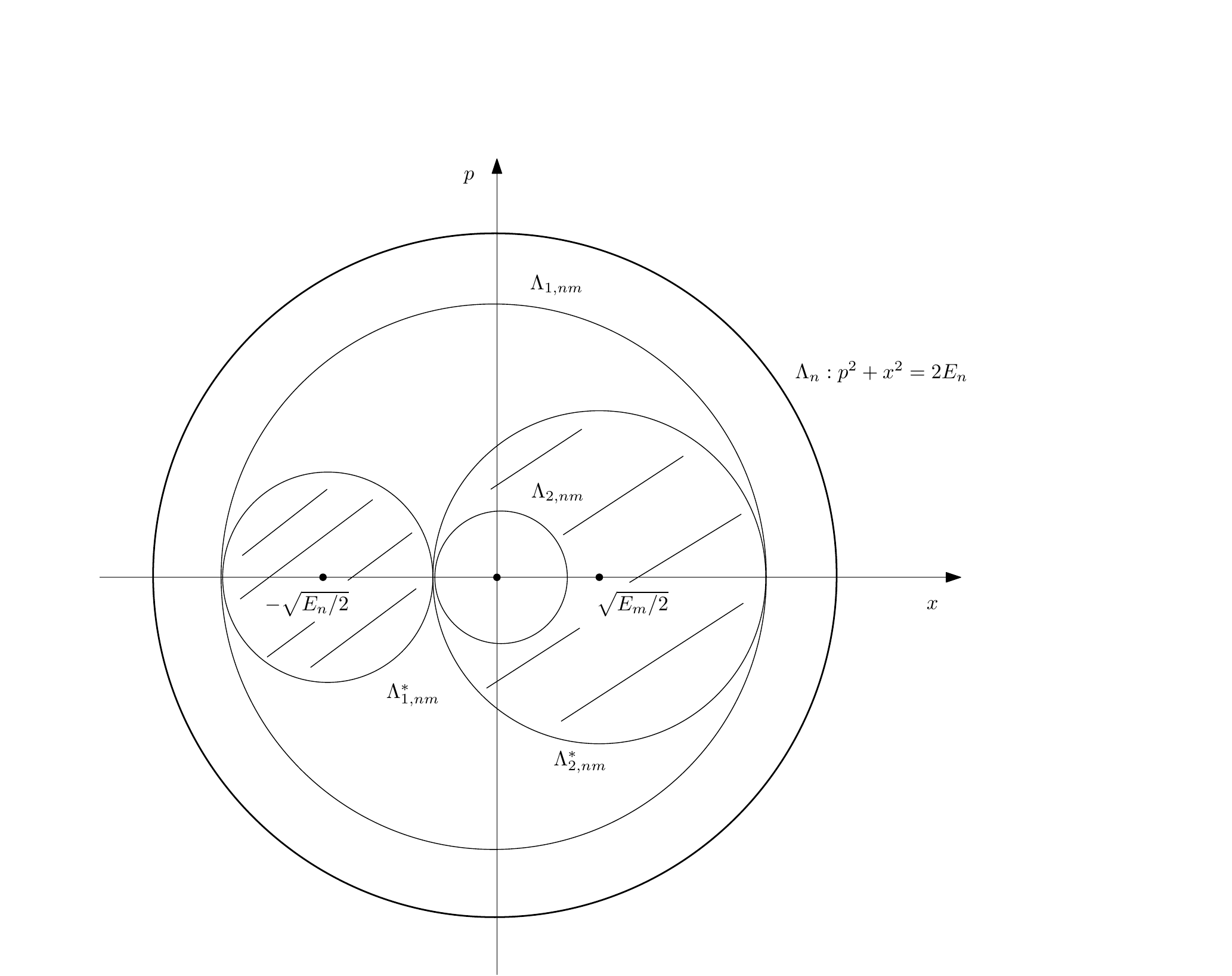}
\caption{{\it Area of existence of stationary points of $F_{\ell,nm}$, $\ell=3,4$  $(\sigma=\si_{1,2} <0)$}}
\label{stationaryp_34nmb}
\end{figure}

The  phases $F_{3,nm}(\sigma,x,p)$ and $F_{4,nm}(\sigma,x,p)$ (eqs. (\ref{phase_3nm}) and  (\ref{phase_4nm}), respectively)  have simple real stationary points in the shaded region of Figures $\ref{stationaryp_34nma}$ , 
$\ref{stationaryp_34nmb}$, and a pair of complex stationary points out of $\Lambda_{n}$. 

By standard stationary phase method, and using involved transformations  which are quite analogous to those applied to the approximation of $\mathcal{W}_{\ell,n}^{\epsilon} \ , \ell=3,4$,  we obtain that 
$\mathcal{W}_{\ell,nm}^{\epsilon} \ , \ell=3,4$, are approximated by the  Airy approximation 
(eq. $(\ref{new_notation_nm})$) .

\subsubsection*{The uniform approximation of $\mathcal{W}^{\hb}_{nm}$}

In Table \ref{table2} we present the main contribution to $\mathcal{W}^{\hb}_{nm}$ for any fixed $(x,p)$ in the strip 
 $\{|x|<\sqrt{2E_{n}^{\hb}} \ , \  p \in \R\}$.

\begin{center}
\begin{tabular}[!hbp]{|c|c|}\hline
region & main contribution to $\mathcal{W}^{\hb}_{nm}(x,p)$ \\
\hline
$p>\sqrt{R^{2}_{nm}-x^2}$ & $\mathcal{W}^{\hb}_{1,nm}\approx\widetilde{\mathcal{W}_{nm}^{\epsilon}}$ \\
\hline
$p\approx\sqrt{R^{2}_{nm}-x^2}$ & $\mathcal{W}^{\hb}_{1,nm}\approx\widetilde{\mathcal{W}_{nm}^{\epsilon}}$ \\
\hline
inside $\Lambda^{*}_{nm}$ & $\mathcal{W}^{\hb}_{3,nm}+\mathcal{W}^{\hb}_{4,nm}\approx\widetilde{\mathcal{W}_{nm}^{\epsilon}}$\\
\hline
$p\approx-\sqrt{R^{2}_{nm}-x^2}$ & $\mathcal{W}^{\hb}_{2,nm}\approx\widetilde{\mathcal{W}_{nm}^{\epsilon}}$ \\
\hline
$p>-\sqrt{R^{2}_{nm}-x^2}$ & $\mathcal{W}^{\hb}_{2,nm}\approx\widetilde{\mathcal{W}_{nm}^{\epsilon}}$ \\
\hline
\end{tabular}
\end{center}
\begin{table}[h!]
\caption{\it{ The main contribution to $\mathcal{W}^{\hb}_{nm}(x,p)$} }
\label{table2}
\end{table}
Therefore,  the leading approximation of $W^{\hb}_{nm}(x,p)$ is 
\begin{eqnarray}\label{approx_w_nm}
\mathcal{W}_{nm}^{\epsilon}(x,p)&\approx& \widetilde{\mathcal{W}_{nm}^{\epsilon}}(x,p) \nonumber\\
&:=& {\pi}^{-1}
e^{-i(n-m)\phi}{\epsilon}^{-2/3}R_{nm}^{-4/3}(R_{nm}^2-\rho_{nm}^2)^{1/3}
Ai\left(\frac{p^2+x^2-R_{nm}^2}{\epsilon^{2/3}{R_{nm}^{4/3}(R_{nm}^2-\rho_{nm}^2)}^{-1/3}}\right)\ , \nonumber \\
\end{eqnarray}
for any $(x,p)$ in the interior of the strip  $\{|x|<\sqrt{2E_{n}^{\hb}} \ , \  p \in \R\}$.

\begin{remark}
The exponential term $e^{-i(n-m)\phi}$ describes the anisotropic exchange of energy between the off-diagonal Wigner eigenfunctions. However, when $(n-m)$ becomes large, it oscillates rapidly, and thus  
$\widetilde{\mathcal{W}_{nm}^{\epsilon}}$ becomes weakly small, which means that the exchange of energy is significant only between neighbouring modes. See also the comments on decoherence after eq.$(\ref{wf_exp_incoh})$ in Section 3.
\end{remark}

\section{Approximate series expansion of the Wigner function}

The proposed approximation of $W^{\epsilon}[u^\epsilon](x,p,t)$ is derived by substituting the approximations (eqs. (\ref{approx_w_n}), (\ref{approx_w_nm}), (\ref{approx_coeff_exp}))
$${\mathcal{W}}^\epsilon_{n}(x,p) \approx \widetilde{\mathcal{W}^{\hb}_{n}}(x,p) \ ,  \mathcal{W}_{nm}^{\epsilon}(x,p)\approx \widetilde{\mathcal{W}_{nm}^{\epsilon}}(x,p)$$
$$c_{nm}^{\hb}\approx \widetilde{\mathcal{C}_{nm}^{\hb}}:=(\mathcal{W}^\hb_{0},\widetilde{\mathcal{W}_{nm}^{\hb}})_{L^2({R}_{xp}^2)} \ ,$$
into the eigenfunction series solution  $W^{\epsilon}[u^\epsilon](x,p,t)$ (eq. $(\ref{wigner_exp})$). Thus  we get the approximation
\begin{equation}\label{wf_exp_new}
W^{\epsilon}[u^\epsilon](x,p,t)\approx\widetilde{W^{\hb}[u^\epsilon]}(x,p,t)= \widetilde{W^{\hb}[u^\epsilon]}_{coh}(x,p) + \widetilde{W^{\hb}[u^\epsilon]}_{incoh}(x,p,t)
\end{equation}
where
\begin{eqnarray}\label{wf_exp_coh_approx}
\widetilde{W^{\hb}[u^\epsilon]}_{coh}(x,p):=\sum_{n=0}^{\infty}\widetilde{\mathcal{C}_{nn}^{\hb}}\,\widetilde{W_{n}^{\hb}}(x,p) 
\end{eqnarray}
and
\begin{equation}\label{wf_exp_incoh_approx}
 \widetilde{W^{\hb}[u^\epsilon]}_{incoh}(x,p,t):=\sum _{n=0}^{\infty}\sum_{m=0, m\ne n}^{\infty} \widetilde{\mathcal{C}_{nm}^{\hb}}\, e^{-\frac{i}{\hb}(E_{n}^{\hb}-E_{m}^{\hb})t}\, \widetilde{W_{nm}^{\hb}}(x,p) \ .
\end{equation}

By using standard WKB estimates of the Schr\"odinger eigenfunctions \cite{Fed}, it is easily shown that the Wigner eigenfunctions  $\wh_{nm}(x,p)$ and the WKB-Wigner eigenfunctions $\mathcal{W}^{\hb}_{nm}(x,p)$ (\ref{wigner_wkb_nm}),  are asymptotically near in the sense that
\begin{eqnarray}\label{wf_exp_near}
\| \wh_{nm}- \mathcal{W}^{\hb}_{nm}\|_{L^{2}(R^{2}_{xp})} =o (\epsilon) \ ,  \ \ \mathrm{as} \ \ \epsilon \to 0 \ .
\end{eqnarray}
However, it is not clear how to derive the $L_2$-asymptotic nearness of $W^{\hb}[u^\epsilon](x,p,t)$ and $\widetilde{W^{\hb}[u^\epsilon](x,p,t)}$, that is 
\begin{eqnarray}\label{wf_exp_near}
\| W^{\hb}[u^\epsilon] - \widetilde{W^{\hb}[u^\epsilon]} \|_{L^{2}(R^{2}_{xp})} =o (\epsilon) \ ,  \ \ \mathrm{as} \ \ \epsilon \to 0 \ , \ t= \mathrm{fixed} \ ,
\end{eqnarray}
although such an estimate is physically anticipated on the basis of energy considerations. The main obstruction is the lack of $L_2$ approximations between the the WKB-Wigner eigenfunctions and their Airy approximations.

Note also that by using polar coordinates $r=(x^2 +p^2)^{1/2}\ ,  \phi=\arctan(p/x)$, we easily get
$$\int\int_{R^{2}_{xp} } \widetilde{W_{nm}^{\hb}[u^\epsilon]}_{incoh}(x,p,t)dxdp = 0 \ ,$$
because $\int_{0}^{2\pi}e^{-\frac{i}{\epsilon} (E^{\hb}_{n}-E^{\hb}_{m})\phi}d\phi =0$. In general,  for anharmonic oscillators
we have the approximation
$$\int\int_{R^{2}_{xp} } \widetilde{W^{\hb}[u^\epsilon]}_{incoh}(x,p,t)dxdp = O(\epsilon) \ .$$

\subsection{Approximation of the expansion coefficients}\label{sec82}

We proceed now to explain how someone can calculate the approximate coefficients (eq. (\ref{approx_coeff_exp}))
$$\widetilde{\mathcal{C}_{nn}^{\hb}}=
({\widetilde{\mathcal{W}}}^{\epsilon}_{0},\widetilde{\mathcal{W}_{nn}^{\hb}})_{L^2({R}_{xp}^2)} \ ,$$
for some particular choices of  the initial amplitude $A_0 (x)$ and phase $S_0 (x)$.

\subsubsection*{{\bf The case of quadratic phase: $S_{0}(x)=\pm x^2/2$ }}

In  this case the semiclassical Wigner function ${\widetilde{\mathcal{W}}}_{0}^\epsilon(x,p)$  (eq.  (\ref{sclwigairy})), is given by
$${\widetilde{\mathcal{W}}}_{0}^\epsilon(x,p)=A^{2}_{0}(x)\delta(p\mp x)  \ . $$
We observe that the different signs in the argument of the Dirac function do not affect the diagonal coefficients 
$\widetilde{\mathcal{C}_{nn}^{\hb}}$, because  the Airy function in 
$\widetilde{\mathcal{W}_{nn}^{\hb}}$ depends on $p^2$, and we get
\begin{equation}\label{coef_cnn_example}
\widetilde{\mathcal{C}_{nn}^{\hb}}=\pi^{-1}\epsilon^{-2/3}(2E^{\hb}_n)^{-1/3}\int_{-\infty}^{+\infty}A^{2}_{0}(x)
Ai \left(\frac{2(x^2-E^{\hb}_n)}{\epsilon^{2/3}{(2E_{n})}^{1/3}}\right)dx \ .
 \end{equation}
 
In the classical limit $\epsilon \to 0$, and for $A_{0}(x)\in C_{0}^{\infty}(R_x) $, by using  the $\mathcal{D'}$- limit
\begin{equation}\label{airy-delta}
\frac{1}{\epsilon}Ai\left(\frac{x}{\epsilon}\right) \rightarrow \delta(x) \ , \ \  \mathrm{as} \ \  \epsilon \to 0 \ , 
\end{equation} 
 and the decomposition formula
 \begin{eqnarray}\label{deltadec}
\delta\left(x^2 -\alpha^2 \right) = \frac{1}{2\alpha}\bigl( \delta\left(x +\alpha\right) + 
\delta\left(x -\alpha\right) \bigr) \ ,
\end{eqnarray}
we derive that $\widetilde{\mathcal{C}_{nn}^{\hb}}$ converges to 
\begin{equation}\label{coef_cnn_example}
\widetilde{\mathcal{C}_{nn}^{0}} \approx \epsilon \frac{1}{4\pi\sqrt{E^{\hb}_{n}}}\left(A^{2}_{0}(\sqrt{E^{\hb}_n})  + A^{2}_{0}(-\sqrt{E^{\hb}_n}) \right) \ .
\end{equation}
Here we assume that $E^{\hb}_{n}$ is fixed,  because is derived by the Bohr-Sommerfeld rule for large $n$ and small $\epsilon$, so that $n\epsilon=$ const. However, if we proceed formally (as tis the way that semiclassical series are used in physical applications), by considering that $\epsilon \ll 1$ and $n$  fixed, we get that 
$\widetilde{\mathcal{C}_{nn}^{0}}=O\left(\epsilon^{1/2}\right)$. Moreover, when $A_{0}(x)$ has compact support, it turns out that only finitely many terms have significant contribution to the coherent part $(\ref{wf_exp_coh_approx})$. Similar approximation holds for $\widetilde{\mathcal{C}_{nm}^{0}}$.

In the special case where $A_{0}(x)\equiv 1$, the quadratic initial phase generates  focal point where the amplitude is infinite, since all the wave energy is periodically concentrated at these points.
Since in this case   the initial datum $u_{0}^{\epsilon} \notin L^{2}(R_x)$, the approximation of the Wigner function,  in principle, cannot be used without further justification. However, we can calculate all integrals analytically by using the formula  (\cite{VS}, eq. (3.93), p. 54)
\begin{eqnarray}\label{aisq_int}
\int_{-\infty}^{+\infty}Ai(z^2-y)dz= 2^{2/3}\pi Ai^{2}\left(  \frac{-y}{2^{2/3}}\right) \ ,
\end{eqnarray}
and we get
\begin{equation}
\widetilde{\mathcal{C}_{nn}^{\hb}}=2^{7/6}\pi \epsilon^{2/3}{(2E^{\hb}_{n})}^{-1/6} Ai^{2}\left(\frac{-(2E^{\hb}_{n})^{2/3}}{ 2^{2/3}\epsilon^{2/3}} \right) \ .
\end{equation}

On the other hand, the different signs of the phase affect the coefficients $\widetilde{\mathcal{C}_{nm}^{\hb}}$ in the incoherent part  through the exponential term $e^{-\frac{i}{\epsilon} (E^{\hb}_{n}-E^{\hb}_{m})\phi(x,p)} \ , \phi(x,p):=\arctan(p/x)$, which for $p=\pm x$ contributes the phase $e^{\mp\frac{i}{\epsilon} (E^{\hb}_{n}-E^{\hb}_{m})\pi/4}$. Therefore, the concentration effects and the formation of focal points, seem to be associated with the incoherent part of the Wigner function. In fact, the dependence on $\epsilon$ of the coefficients of the coherent part of the Wigner function implies that the coherent amplitude is bounded for $\epsilon \ll 1$.

In the case of Gaussian amplitude  $A_{0}(x)= e^{{-x^2}/2} $, we have $u_{0}^{\epsilon} \in L^{2}(R_x)$, and our approximation can be applied.
For the approximate computation of the coefficients we use use the asymptotic decomposition formula \cite{GMa}
\begin{eqnarray}\label{airydec}
\frac{1}{\epsilon}Ai\left( \frac{x^2 -\alpha^2}{\epsilon}  \right)  \nonumber 
\asymp \frac{1}{2\alpha}\left[\frac{1}{\epsilon}Ai\left( \frac{x +\alpha}{\epsilon}  \right) +\frac{1}{\epsilon}Ai\left( \frac{x -\alpha}{\epsilon}  \right) \right] \ , \ \ 
\mathrm{as} \ \  \epsilon \ll 1 \ .
\end{eqnarray}
The symbol $\asymp$ means that we omit terms of the form $(\epsilon^{3/2})^{k}\sin\left(\epsilon^{-3/2} \frac23 \alpha^3\right)\delta^{(\ell)}(x)$, $k \ , \ell = 1\ ,  2 \ , \dots$.
Then, we have
\begin{eqnarray}\label{sumgaussian}
\widetilde{\mathcal{C}_{nn}^{\hb}} \approx \frac {1}{2a}\left[\int_{R_x}^{}e^{-x^2}Ai\left(\frac{x+a}{b}\right)\, dx+
\int_{R_x}^{} e^{-x^2}Ai\left(\frac{x-a}{b}\right)\, dx \right]
\end{eqnarray}
where $a:=\sqrt{E_{n}^{\hb}}$ and $b:=2^{-1}\hb^{2/3}{(2E_{n}^{\hb})}^{1/3}$. 

The integrals into $(\ref{sumgaussian})$ are Airy transforms of the Gaussian function, and they can calculated by the formula ( \cite{VS}, p. 78, eq. (4.31)),
\begin{eqnarray}
\phi_{\alpha}(x)&=&\frac{1}{|\alpha|}\int_{R}^{}e^{-y^2}Ai\left(\frac{x-y}{\alpha}\right)\, dy \nonumber\\
&=&\frac{\sqrt{\pi}}{|\alpha|}e^{\left(x+\frac{1}{24\alpha^3}\right)/4\alpha^3}Ai\left(\frac{x}{\alpha}+\frac{1}{16\alpha^4}\right) \ , \ \ \alpha \in R \ .
\end{eqnarray}
Thus,  we  approximate  $\widetilde{\mathcal{C}_{nn}^{\hb}}$ by
\begin{eqnarray}
\widetilde{\mathcal{C}_{nn}^{\hb}} \nonumber
\approx \frac{\sqrt{\pi}e^{\frac{1}{96 b^6}}}{2ab}\left(e^{a/4b^3} Ai\left(\frac{a}{b}+\frac{1}{16b^4}\right) +e^{-a/4b^3} Ai\left(\frac{-a}{b}+\frac{1}{16b^4}\right) \right) \ ,
\end{eqnarray}
with $a:=\sqrt{E_{n}^{\hb}}$ and $b:=2^{-1}\hb^{2/3}{(2E_{n}^{\hb})}^{1/3}$.

\subsubsection*{{\bf The case of cubic phase: $S_{0}'''(x) \neq 0$.}}

The simplest initial phase having this property is $S_{0}(x)=-x^3/6$ . Such initial phase generates a cusp caustic. 
In this case, the semiclassical Wigner function ${\widetilde{\mathcal{W}}}_{0}^\epsilon$  (eq. (\ref{sclwigairy})), is given by
$${\widetilde{\mathcal{W}}}_{0}^\epsilon(x,p)= \frac{2}{\epsilon^{2/3}} A_{0}^{2}(x)Ai\left(\frac{2}{\epsilon^{2/3}}\left(p-\frac{x^2}{2}\right)\right) $$


Now the coefficients $\widetilde{\mathcal{C}_{nn}^{\hb}}$ are written in terms of the semiclassical Wigner function of the
\begin{eqnarray}\label{approx_cnm}
\widetilde{\mathcal{C}_{nn}^{\hb}}=\int_{R_{x}}A_{0}^{2}(x) Q^{\hb}_{n}(x)dx  \ ,
\end{eqnarray}
where
\begin{eqnarray}\label{int_qn}
\frac{(2E^{\hb}_n)^{1/3}}{\pi}{\hb^{4/3}}Q^{\hb}_{n}(x):=\int_{R_{p}} Ai\left(\frac{2}{\epsilon^{2/3}}\left(p+\frac{x^2}{2}\right)\right) Ai\left(\frac{p^2+x^2-2E^{\hb}_n}{\epsilon^{2/3}{(2E^{\hb}_{n})}^{1/3}}\right)dp \ . \nonumber
\\
\end{eqnarray}
By using the decomposition formula $(\ref{airydec})$, we approximately decompose $Q^{\epsilon}_{n}$ as
\begin{equation}
Q^{\hb}_{n}(x) \approx \frac{1}{\sqrt{2E^{\hb}_n -x^2}}\left[ Q^{\hb}_{n+}(x)+ Q^{\hb }_{n-}(x)\right] \ ,
\end{equation}
where
\begin{eqnarray}\label{int_qnpm}
\frac{(2E^{\hb}_n)^{1/3}}{\pi}{\hb^{4/3}}Q^{\hb}_{n\pm}(x):=\int_{R_{p}} Ai\left(\frac{2}{\epsilon^{2/3}}\left(p+\frac{x^2}{2}\right)\right) Ai\left( \frac{p \pm \sqrt{2E^{\hb}_n -x^2}}{\epsilon^{2/3}{(2E^{\hb}_{n})}^{1/6}}\right)dp \ . \nonumber \\
\nonumber\\
\end{eqnarray}

The integrals $Q^{\hb}_{n\pm}$ are calculated by using the formula (\cite{VS}, eq. (3.108), p. 57)
\[\frac{1}{\mid \alpha \beta\mid}\int_{-\infty}^{\infty}Ai\left(\frac{z+a}{\alpha}\right)Ai\left(\frac{z+b}{\beta}\right)dz =
\left\{\begin{array}{lr}
\delta(b-a) \ \ \hfill  \mathrm{if} \ \   \beta=\alpha \\
\frac{1}{|\beta^3 -\alpha^3|^{1/3}}Ai\left( \frac{b-a}{ (\beta^3 -\alpha^3)^{1/3}}\right) \ \ \ \ \mathrm{if} \ \ \beta \neq \alpha \ .
\end{array} \right.
\]
The integration leads to
\begin{eqnarray}\label{qnpm}
Q^{\hb}_{n\pm}(x)= \frac{\pi}{(2E^{\hb}_n)^{1/6}}
\frac{1}{\hb^{2/3}}\frac{(2E^{\hb}_n)^{1/6}  }
 {\mid (2E^{\hb}_n)^{1/2}- \frac18 \mid^{1/3}}\, \nonumber \\
 \times
Ai\left( \frac{1}{\epsilon^{2/3}} \frac{ -\frac{x^2}{2}\pm  \sqrt{2E^{\hb}_n -x^2}}{\mid (2E^{\hb}_n)^{1/2}  -\frac18  \mid^{1/3}}  \right) \ .
\end{eqnarray}

It is important to observe that these approximation formulas are smooth and they don't posses any singularity at the turning points.  This smoothness is a consequence of the  {\it uniformization procedure} for the construction of the approximate Wigner eigenfunctions. In fact, such smooth approximations of the coefficients cannot be derived  directly in the configuration space by using  (\ref{sercoef}),  because the Schr\"odinger eigenfunctions are weakly singular at the turning points.

We also observe that the argument of the Airy function in 
$Q^{\hb}_{n-}(x)$ is always negative, while that in $Q^{\hb}_{n+}(x)$ vanishes when  $x^2=2\sqrt{2E^{\hb}_n -x^2}$. Therefore, for small $\epsilon$, we expect that  the main contribution to the integral  $(\ref{approx_cnm})$ comes from $Q^{\hb}_{n+}(x)$ due to concentration of the Airy function, while the contribution of $Q^{\hb}_{n-}(x)$ is expected to me negligible due to fast oscillations of the Airy function.

\subsection{Approximation of the energy density}

The approximate solution $(\ref{wf_exp_new})$  of the Wigner equation, implies an approximation of the amplitude of the wavefunction, and, more precisely, a decomposition  into a coherent and an incoherent component of the of the energy density \footnote {This decomposition provides, in principle, a way to study large-time asymptotics of the transport and the Hamilton-Jacobi equations.} . 

Let $u^\epsilon(x,t)=\alpha^\epsilon (x,t) e^{i\phi^\epsilon (x,t)}$ be the polar decomposition of the wavefunction. 
By integrating  $(\ref{wf_exp_new})$ with respect to the  momentum $p$, for some fixed $(x,t)$, we  get the following approximate decomposition of the energy density $\eta^{\epsilon} (x,t)= \mid\alpha^\epsilon (x,t)\mid^{2}$
\begin{equation}\label{approx_ampl}
\eta^{\epsilon} (x,t)= \eta^{\epsilon}_{coh} (x,t) + \eta^{\epsilon}_{incoh} (x,t) \ ,
\end{equation}
where
\begin{equation}\label{approx_coh_ampl}
\eta^{\epsilon}_{coh} (x,t) =\int_{R_p}W^{\hb}[u^\epsilon]_{coh}(x,p)dp \approx
\sum_{n=0}^{\infty}\widetilde{\mathcal{C}_{nn}^{\hb}} \int_{R_p} \widetilde{W_{nn}^{\hb}}(x,p) dp  
\end{equation}
and 
\begin{eqnarray}\label{approx_incoh_ampl}
\eta^{\epsilon}_{incoh} (x,t)&=&\int_{R_p}W^{\hb}[u^\epsilon]_{coh}(x,p,t)dp \nonumber \\
&\approx&\sum_{n=0}^{\infty}\sum_{m=0 \ , m\ne n}^{\infty}\widetilde{\mathcal{C}_{nm}^{\hb}}\, e^{-\frac{i}{\hb}(E_{n}^{\hb}-E_{m}^{\hb})t}\, \int_{R_p} \widetilde{W_{nm}^{\hb}}(x,p) dp\ .
\end{eqnarray}

The integrals $\int_{R_p} \widetilde{W_{nn}^{\hb}}(x,p) dp$  in equation $(\ref{approx_coh_ampl})$, can be  calculated by using the formula $(\ref{aisq_int})$.
Thus, we obtain the following approximation of the coherent component of the intensity
\begin{eqnarray}\label{coh_ampl_approx}
\eta^{\epsilon}_{coh} (x)&\approx & \sum_{n=0}^{\infty}\widetilde{\mathcal{C}_{nn}^{\hb}} \int_{R_p} \widetilde{W_{nn}^{\hb}}(x,p) dp\nonumber\\
&=&\sum_{n=0}^{\infty}\widetilde{\mathcal{C}_{nn}^{\hb}}\, 2^{2/3}\epsilon^{-1/3}(2E_{n}^{\hb})^{-1/6}Ai^{2}\left(\frac{-(2E^{\hb}_n -x^2)}{2^{2/3}\epsilon^{2/3}(2E_{n}^{\hb})^{1/3}}\right) \ .
\end{eqnarray}
Note that $ \eta^{\epsilon}_{coh} (x,t) $ is always a positive quantity, since $\widetilde{\mathcal{C}_{nn}^{\hb}}> 0$ by their definition and the construction of the approximation. On the other hand,
$\eta^{\epsilon}_{incoh} (x,t)$ oscillates and changes sign as time evolves. Since as the time $t$ increases, the exponential terms $e^{-\frac{i}{\hb}(E_{n}^{\hb}-E_{m}^{\hb})t}$ tend weakly to zero, we expect that the incoherent part is weakly negligible for large time.

For the incoherent component of the density we have 
\begin{eqnarray}\label{incoh_ampl_approx}
&&\eta^{\epsilon}_{incoh} (x,t) \approx\nonumber\\
&&
\sum_{n=0}^{\infty}\sum_{m=0 \ , m\ne n}^{\infty}\widetilde{ c_{nm}^{\hb}}(0)\, e^{-\frac{i}{\hb}(E_{n}^{\hb}-E_{m}^{\hb})t}\,{\pi}^{-1}{\epsilon}^{-2/3}R_{nm}^{-4/3}(R_{nm}^2-\rho_{nm}^2)^{1/3} \nonumber\\
&&\times \int_{-\infty}^{\infty}
e^{-\frac{i}{\epsilon} (E^{\hb}_{n}-E^{\hb}_{m})\arctan(p/x)}
Ai\left(\frac{x^2+p^2-R_{nm}^2}{\epsilon^{2/3}{R_{nm}^{4/3}(R_{nm}^2-\rho_{nm}^2)}^{-1/3}}\right) dp \ .
\end{eqnarray}
The integrals in $(\ref{incoh_ampl_approx})$ cannot calculated analytically at a general space-time point $(x \ ,t)$. Nevertheless, it can be shown, by using the Riemann-Lebesgue lemma  and the exponential decay of the Airy function for large positive argument, that they are convergent. At the special position $x=0$, the exponential term  
disappears. Then we can calculate the integrals by using the formula $(\ref{aisq_int})$, and we get

\begin{eqnarray}\label{incoh_ampl_approx0}
&&\eta^{\epsilon}_{incoh} (x=0,t) \approx \nonumber 
\sum_{n=0}^{\infty}\sum_{m=0 \ , m\ne n}^{\infty}\widetilde{ c_{nm}^{\hb}}(0)\, e^{-\frac{i}{\hb}(E_{n}^{\hb}-E_{m}^{\hb})t}\nonumber\\
&&\times 2^{2/3}
{\epsilon}^{-1/3}R_{nm}^{-2/3}(R_{nm}^2-\rho_{nm}^2)^{1/6} 
Ai^{2}\left(\frac{R_{nm}^{2}}{{\epsilon}^{2/3}R_{nm}^{4/3}(R_{nm}^2-\rho_{nm}^2)^{-1/3}}\right) \ .
\end{eqnarray}

At this point we can make the following important remark. In geometrical optics, focal points and caustic formation appear along space-time curves.
Therefore, from the fact that the coherent part of the amplitude is time-independent, we expect  that generation of effects is associated with the incoherent part. This implies, in turn,  that possible amplification of the wave intensity, as $\epsilon$ diminishes, should be encoded in the off-diagonal coefficients.

The detailed investigation of  the formation of the singularities as the time evolves, is an open important problem which requires the calculation or the approximation of the integrals in 
$(\ref{incoh_ampl_approx})$. A preparatory step would be  the systematic study of the trigonometric series
$(\ref{incoh_ampl_approx0})$ which is still very complicated due to the Airy functions appearing in its coefficients.

\appendix
\numberwithin{equation}{section}
\section*{Appendices}
\section{QM in configuration space: Schr\"odinger equation}

\subsection{Eigenfunction series expansion of the wave function}\label{32}

We assume that  the potential $V(x) \in C^{\infty}(R_x)$ is positive, real valued,  and  $\lim_{| x| \rightarrow \infty}V(x)=\infty$. For each fixed $\epsilon \in(0,\epsilon_0)$, with arbitrary $\epsilon_0>0$,  the spectral problem 
\begin{eqnarray}\label{statschr}
\widehat{H}^{\epsilon}v^{\epsilon}(x)=\left[-\frac{\epsilon^{2}}{2}\frac{d^{2}}{dx^{2}}+V(x)\right]v^{\epsilon}(x)=E^{\epsilon}v^{\epsilon}(x) \quad .
\end{eqnarray}
has purely discrete spectrum with eigenvalues arranged in an increasing sequence,
\begin{eqnarray}\label{energy_seq}
0<E_{0}^{\epsilon}< E_{1}^{\epsilon}\leq\ldots\leq E_{n}^{\epsilon}\leq\ldots, \quad \lim _{n\rightarrow \infty}E_{n}^{\epsilon}=+\infty \quad  \ .
\end{eqnarray}
 The corresponding eigenfunctions $v_{n}^{\epsilon}(x) \in L^2(R_x)$ form an orthonormal set with respect to the $L^{2}-$ inner product (see e.g. \cite{BS, HS}).

Then, by separation of variables (Fourier method),
the solution $u^{\epsilon}(x,t)$ of the Cauchy problem 
\begin{eqnarray}
i\epsilon{\partial_t}u^{\epsilon}(x,t)&=&\left[-\frac{\epsilon ^2}{2}\partial_{xx}+V(x)\right]u^{\epsilon}(x,t)\quad , \quad x\in R_{x} \quad ,\, \, \, t\in [0,T) \ , \label{schr_eq_1} \\
u^{\epsilon}(x,t=0)&=&u_{0}^{\epsilon}(x)=A_{0}(x)\, e^{\frac{i}{\epsilon}S_{0}(x)} \ . \label{initialdata}
\end{eqnarray}
with initial data  $A_{0}(x)\in C_{0}^{\infty}(R_x)$ and $S_{0}(x) \in C^{\infty}(R_x)$, and
for some positive constant $T< \infty$,  is given by the eigenfunction series 
\begin{eqnarray}\label{serexpsch}
u^{\epsilon}(x,t)=\sum_{n=0}^{\infty}c^{\epsilon}_{n}v_{n}^{\epsilon}(x)e^{-\frac{i}{\epsilon}E_{n}^{\epsilon} t} \ ,
\end{eqnarray}
where coefficients  
$c^{\epsilon}_{n}$ are the $L^{2}-$projections of the initial data (\ref{initialdata}) onto the eigenfunctions
\begin{eqnarray}\label{sercoef}
c^{\epsilon}_{n}(0)=(u^{\epsilon}_{0},v_{n}^{\epsilon})_{L^{2}(R)}\quad \mathrm{for \,\,\,  all}\,\, \, n=0,1,2,\ldots \ \ .
\end{eqnarray}

The solution $u^{\epsilon}(x,t)$ belongs to  $L^{2}(R_x)$ and it conserves the quantum energy, that is  
$\Vert u^{\hb}(x,t)\Vert_{L^{2}(R_{x})}= \Vert u_{0}^{\epsilon}(x)\Vert_{L^{2}(R_{x})}$, for any fixed $t \in [0,T)$.

Although the eigenfunction expansion of the wavefunction is an exact 
solution of the Cauchy problem (\ref{schr_eq_1})-(\ref{initialdata}), it is, in general, very slowly convergent for small values of the semiclassical parameter $\epsilon$. In such cases asymptotic approximations of the eigenvalues and eigenfunctions  are constructed by the WKB method, but these approximate eigenfunctions have singularities on the turning points where their amplitude blows up. 
\subsection{WKB asymptotic expansion of eigenfunctions}

According to the WKB method,  the eigenvalues $E_{n}^{\hb}$  are approximated for small $\epsilon$, by the {\it Bohr-Sommerfeld quantization rule} (see, e.g., Fedoriuk \cite{Fed},Yosida \cite{Y}).
Assuming that the potential has the shape of a single well, and $x_1$, $x_2$ are the turning points satisfying $V(x_1(E_{n}^{\hb}))=V(x_1(E_{n}^{\hb}))=E_{n}^{\hb}$, with $x_1(E_{n}^{\hb})<x_2(E_{n}^{\hb})$, this rule implies that  the eigenvalues $E_{n}^{\hb}$ are approximate solutions of the equation
\begin{eqnarray}\label{bohrsom}
f(E_{n}^{\hb})\equiv\int_{x_{1}(E_{n}^{\hb})}^{x_{2}(E_{n}^{\hb})}\sqrt{2(E_{n}^{\hb}-V(x))}\, dx\approx\pi\left(n+\frac{1}{2}\right)\epsilon \ ,
\end{eqnarray}
for large $n$, so that  $n\epsilon$ being constant. 
\begin{figure}[h]
\centering
\includegraphics[width=0.7\textwidth]{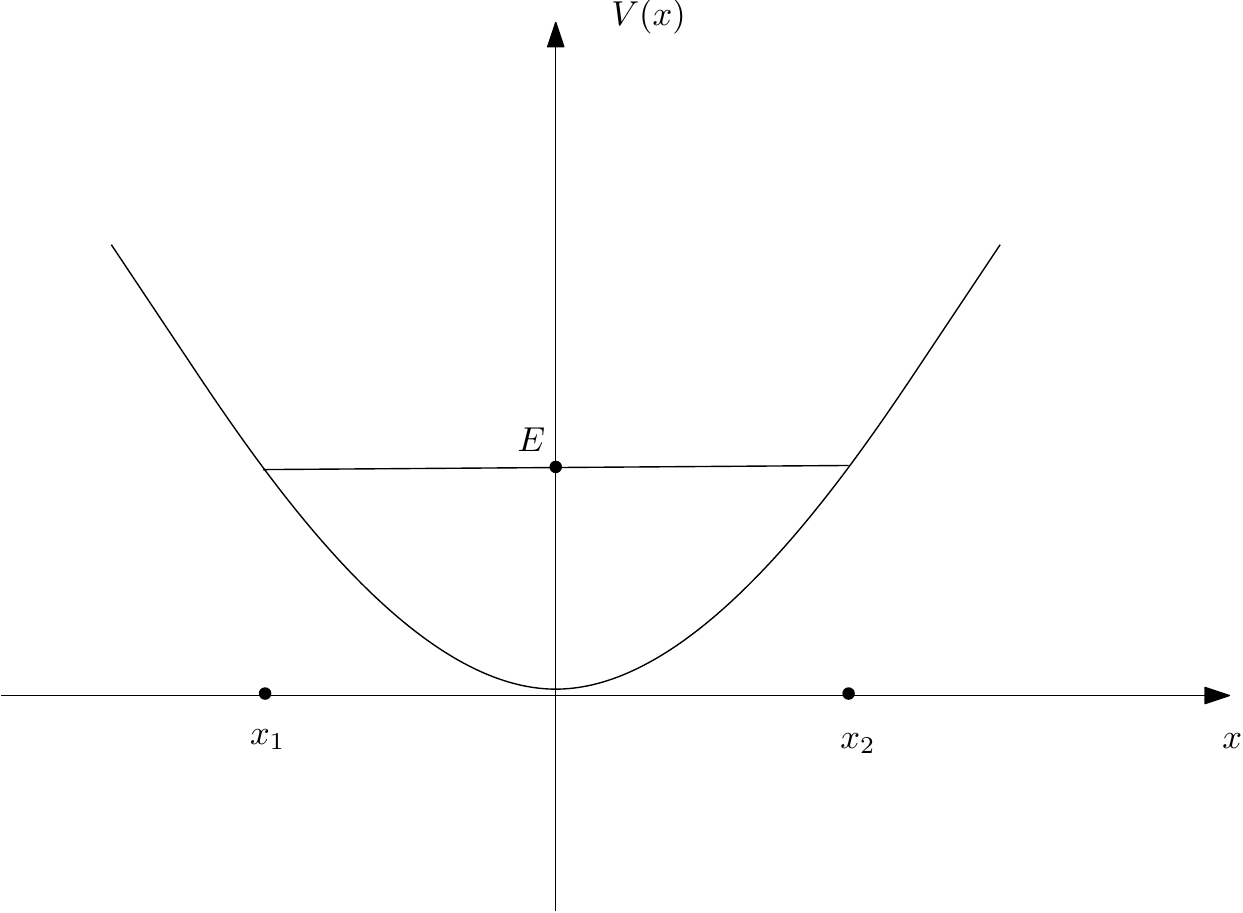}
\caption{{\it Single well potential}}
\label{well}
\end{figure}


The asymptotic approximations of the corresponding eigenfunctions, are given by
\begin{itemize}
\item $x>x_{2}(E_{n}^{\hb})$ 
\end{itemize}
\begin{eqnarray}\label{wkbout1}
v_{n}^{\epsilon}(x)\approx \psi_{n}^{\epsilon}(x)={[2(V(x)-E_{n}^{\hb})]}^{-1/4}\exp\left(-\frac{1}{\epsilon} \int_{x_{2}(E_{n}^{\hb})}^{x}\sqrt{2(V(t)-E_{n}^{\hb})} \, dt\right)\quad  \ ,\nonumber\\
&&
\end{eqnarray}
\begin{itemize}
\item
 $x_{1}(E_{n}^{\hb})<x<x_{2}(E_{n}^{\hb})$ 
 \end{itemize}
\begin{eqnarray}\label{wkbsolution_1}
v_{n}^{\epsilon}(x)\approx\psi_{n}^{\epsilon}(x)={[2(E_{n}^{\hb}-V(x))]}^{-1/4}\cos\left({\frac{1}{\epsilon}\int_{x_{2}(E_{n}^{\hb})}^{x}\sqrt{2(E_{n}^{\hb}-V(t))} \, dt+\frac{\pi}{4}}\right)\quad \ , \nonumber \\
&& \,
\end{eqnarray}
\begin{itemize}
\item
 $x<x_{1}(E_{n}^{\hb})$ 
\end{itemize}
\begin{eqnarray}\label{wkbout2}
v_{n}^{\epsilon}(x)\approx\psi_{n}^{\epsilon}(x)= (-1)^n{[2(V(x)-E_{n}^{\hb})]}^{-1/4}\exp\left(\frac{1}{\epsilon}\int_{x_{2}(E_{n}^{\hb})}^{x}\sqrt{2(V(t)-E_{n}^{\hb})} \, dt\right)\quad \ . 
\end{eqnarray}

Clearly, these approximations break down at the turning points, where $V(x)=E_{n}^{\hb}$, since the amplitudes $[2(V(x)-E_{n}^{\hb})]^{-1/4}$ diverge at these points.
In the  classically allowed region $x_{1}(E_{n}^{\hb})<x<x_{2}(E_{n}^{\hb})$ the eigenfunction is rapidly oscillatory and in the classically forbidden regions $x>x_{2}(E_{n}^{\hb})$ and $x<x_{1}(E_{n}^{\hb})$ is exponentially decaying.

In the case of the harmonic oscillator $V(x)=x^{2}/2$, the integral in (\ref{bohrsom}) is computed analytically, and we get the approximation
\begin{eqnarray}\label{energy_appr}
E_{n}^\epsilon\approx\left(n+\frac{1}{2}\right)\epsilon \ .
\end{eqnarray}

In the oscillatory region $E_{n}^{\hb}>x^2/2$,  the approximation (\ref{wkbsolution_1}) is written in the form
\begin{equation}\label{wkbcos}
v_{n}^{\epsilon}(x)\approx \left(\frac{2}{\pi}\right)^{1/2}\psi_{n}^{\epsilon}(x)
=\left(\frac{2}{\pi}\right)^{1/2}\left(2E_{n}^{\hb}-x^2\right)^{-1/4}\cos\left({\frac{1}{\epsilon}\int_{\sqrt{2E_{n}^{\hb}}}^{x}\sqrt{2E_{n}^{\hb}-t^2} \, dt+\frac{\pi}{4}}\right)\ .
\end{equation}
In order to emphasize the {\it two-phase structure of the WKB approximations near the turning points}, we rewrite (\ref{wkbcos}) in terms of complex exponentials
\begin{equation}\label{wkbhosc}
v_{n}^{\hb}(x)\approx\left(\frac{2}{\pi}\right)^{1/2}\psi^{\epsilon}_{n}(x)={A_{n}^{\hb}}^{+}(x)\,
e^{\frac{i}{\epsilon}{S_{n}^{\hb}}^{+}(x)}+{A_{n}^{\hb}}^{-}(x)\,
e^{\frac{i}{\epsilon}{S_{n}^\hb}^{-}(x)} \ .
\end{equation}
where the amplitudes ${A_{n}^{\hb}}^{\pm}$ and the phases ${S_{n}^\hb}^{\pm}$ are given by the formulae
\begin{equation}
{A_{n}^{\hb}}^{\pm}(x):=\frac{1}{2}\left(\frac{2}{\pi}\right)^{1/2}\left(2E_{n}^{\hb}-x^2\right)^{-1/4}e^{\pm i\pi /4}\quad ,\label{wkbampl_1}\\
\end{equation}
\begin{equation}\label{wkbph_1}
{S_{n}^{\hb}}^{+}(x)=-{S_{n}^{\hb}}^{-}(x)=\int_{\sqrt{2E_{n}^{\hb}}}^{x}\sqrt{2E_{n}^\epsilon - t^2}dt\quad .
\end{equation}
The amplitudes $A_{n}^{\pm}(x)$ diverge at the turning points $x=\pm\sqrt{2E_{n}^{\hb}}$ (caustics) and the exponentials
$e^{\pm i\pi /4}$ take care of the phase shift there.

The approximation (\ref{energy_appr}) of the eigenvalues coincide with the exact eigenvalues 
$E_{n}^{\hb} = \left(n +1/ 2\right)\hb $,  $n=0,1,\ldots$,
while the exact eigenfunctions are given by (see, e.g. \cite{Foc, Tak})
\begin{equation}\label{eigenf_harm}
\vh_{n}(x)=\frac{e^{-x^2/2\epsilon}}{(\pi\epsilon)^{1/4}\sqrt{2^{n}n!}}\, H_{n}\left(\frac{x}{\sqrt{\epsilon}}\right)\, ,\quad \mathrm{for}\quad n=0,1,\ldots \ ,
\end{equation}
where $H_n(x):=(-1)^n e^{x^2}\frac{d^n}{dx^n}\left(e^{-x^2}\right)$ are the Hermite polynomials.

By exploiting  appropriate asymptotic expansions of the Hermite polynomials \cite{Do}, we can check that  the asymptotic expansion of (\ref{eigenf_harm}) coincides, in the leading order,  with the WKB approximation  (\ref{wkbcos}) in the  oscillatory region.


\section{Stationary phase formulae}\label{app_a}
\subsection{The case of a simple stationary point}

We consider the integral
\begin{equation}\label{integral}
I(\lambda)=\int_{a}^{b}f(x)\, e^{i\lambda\phi(x)} \,  dx
\end{equation}
where $f\in C[a,b]$, and $\phi\in C^2[a,b]$ is a real-valued function with a simple stationary point 
$x=c \in (a,b)$ such that $\phi'(c)=0$ and $\phi''(c)\not =0$. 
Then, the following approximation formula holds 
\begin{equation}\label{statphform}
I(\lambda)= e^{i\lambda\phi(c)+i\delta
\pi/4}f(c)\left[\frac{2\pi}{\lambda|\phi''(c)|}\right]^{1/2} +O(\lambda^{-3/2}) \ , \ \ \ \lambda\rightarrow \infty \ ,
\end{equation}
with $\delta=\sgn\phi''(c)$ (see e.g. \cite{BH}, Ch. 6, or \cite{Bor}, Ch. 2). 


\subsection{The case of  two coalescing stationary points (uniform formula)}

We consider the integral
$$I(\lambda ,\alpha)=\int_{-\infty}^{\infty} f(x)\,  e^{i\lambda \phi(x,\alpha)}dx, $$
where the phase depends on the parameter $\alpha>0$, and look for the asymptotic behavior of $I$, as
$\lambda \rightarrow \infty$. We assume again that $f\in C[a,b]$, and that the phase function $\phi\in
C^{\infty}$ has two stationary points, $x_1(\alpha)$ and
$x_2(\alpha),$ which approach the same limit $x_0$ when
$\alpha\rightarrow 0.$ Let $\phi_{xx}(x_1,\alpha)<0$ and
$\phi_{xx}(x_2,\alpha)>0$.
Then, the approximation
\begin{eqnarray}\label{ap4}
I(\lambda ,\alpha)= e^{i\lambda \phi_0(\alpha)}\left[2\pi
A_{0}(\alpha)\lambda ^{-1/3}Ai(-\lambda ^{2/3}\xi)-2\pi i
B_{0}(\alpha)\lambda ^{-2/3} Ai'(-\lambda ^{2/3}\xi)+C(\lambda
,\xi)\right] \ ,
\end{eqnarray}
with
\begin{eqnarray}\label{ap23}
\phi_0(\alpha)=\frac{1}{2}\left(\phi(x_1(\alpha),\alpha)+\phi(x_2(\alpha),\alpha)\right) \ ,
\end{eqnarray}

\begin{eqnarray}\label{ap13}
\xi(\alpha)=\left[\
\frac{3}{4}\left(\phi(x_1(\alpha),\alpha)-\phi(x_2(\alpha),\alpha)\right)\
\right]^{2/3} \ ,
\end{eqnarray}
and
\begin{eqnarray}\label{A0}
A_0=2^{-1/2}\xi^{1/4}\left[\frac{f(x_2)}{\sqrt{\phi_{xx}(x_2,\alpha)}}+\frac{f(x_1)}{\sqrt{|\phi_{xx}(x_1,\alpha)|}}\right] \ ,
\end{eqnarray}
\begin{eqnarray}\label{B0}
B_0=-2^{-1/2}\xi^{-1/4}\left[\frac{f(x_1)}{\sqrt{|\phi_{xx}(x_1,\alpha)|}}-\frac{f(x_2)}{\sqrt{\phi_{xx}(x_2,\alpha)}}\right] \ ,
\end{eqnarray}
holds uniformly for any  $\alpha >0$. 

As $\alpha\rightarrow 0^+$ we use the following approximations,

\begin{eqnarray}\label{app10}
\phi_{0}(\alpha)\approx \phi(0,0) \ ,
\end{eqnarray}

\begin{eqnarray}\label{32}
\xi^{1/4}\approx \left[-\partial_{x\alpha}\phi\
\left(\frac{\partial_{xxx}\phi}{2}\right)^{-1/3}\alpha\right]^{1/4} \ ,
\end{eqnarray}
and 
\begin{eqnarray}\label{30}
\mid \partial_{xx}\phi(x_1(\alpha),\alpha)\mid \ \approx (-2\partial_{xxx}\phi\
\partial_{x\alpha}\phi\ \alpha)^{1/2}\, ,
\end{eqnarray}
\begin{eqnarray}\label{31}
\mid \partial_{xx}\phi(x_2(\alpha),\alpha)\mid \ \approx (-2\partial_{xxx}\phi\
\partial_{x\alpha}\phi\ \alpha)^{1/2} \ ,
\end{eqnarray}
where the derivatives are calculated at the point $(x, \alpha=0)$.

The approximation (\ref{ap4}) has been constructed by Chester, Friedman and Ursell in \cite{CFU} for the case of analytic phases,  and a concise derivation is presented in \cite{Bor}.


\section{Berry's semiclassical Wigner function}

We consider the Wigner transform
\begin{eqnarray}\label{rescaled_wig}
{\mathcal{W}}^{\epsilon}(x,p)=\frac{1}{\pi\epsilon}\int_{R}^{}\ \psi^{\epsilon}(x+\si)\,
\overline{{\psi}^{\epsilon}}(x-\si)\, \ e^{-\frac{i}{\epsilon}2p\si}\, d\si
\end{eqnarray}
of the WKB wave function
\begin{equation}\label{wkb_1}
\psi^{\epsilon}(x)=A(x)\, e^{iS(x)/\epsilon} \ ,
\end{equation}
where the amplitude $A$ and the phase $S$ are smooth, real-valued functions, and that $S^{'}(x)$ is globally convex.

We write (\ref{rescaled_wig})  in the form of Fourier integral 
\begin{equation}\label{wigwkb}
{\mathcal{W}}^{\epsilon}(x,p)=\frac{1}{\pi\epsilon}\int_{R}^{}D(\si,x)\,
e^{i\frac{1}{\epsilon}F(\si,x,p)}\, d\si \quad ,
\end{equation}
where
\begin{equation}
D(\si,x)=A(x+\si)A(x-\si)
\end{equation}
is the Wigner amplitude, and
\begin{equation}\label{wigphase}
F(\si,x,p)=S(x+\si)-S(x-\si)-2p\si
\end{equation}
is the Wigner phase. 

For any fixed $(x,p)$, the critical  points of the
phase $F(\si,x,p)$ are the roots of
\begin{equation}\label{alpha}
\partial_{\si}F(\si,x,p)=S^{'}(x+\si)+S^{'}(x-\si)-2p=0 \ .
\end{equation}

Assuming that $S^{'''}(x)\not=0$, we have 
\begin{equation}
\partial_{\si\si}F(\si=0,x,p)=0\, , \quad \partial_{\si\si\si}F(\si=0,x,p)=2S^{'''}(x)\not=0 \ ,
\end{equation}
and therefore $\si=0$ is a double stationary point of $F$.

\begin{figure}[h]
\centering
\includegraphics[width=0.7\textwidth]{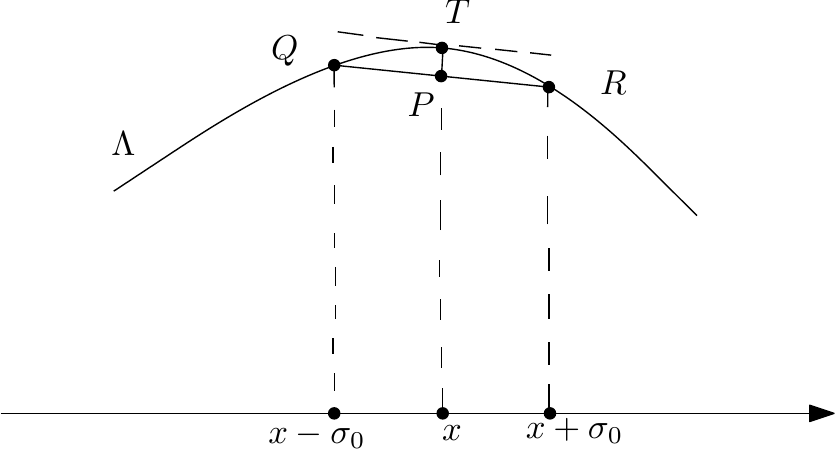}
\caption{{\it Lagrangian curve \& Berry's chord}}
\label{berry_ch}
\end{figure}

Berry \cite{Ber} has introduced  an invariant geometrical interpretation of this equation (Figure \ref{berry_ch}), by observing that 
(\ref{alpha}) has a pair of symmetric roots $\pm\si_{0}(x,p)$ such that the point $P=(x,p)$ be the middle of a chord $QR$ (Berry's chord) with endpoints $Q(x-\si_{0} \ , S^{'}(x-\si_{0}))$ and $R(x-\si_{0}  \ , S^{'}(x+\si_{0}))$
on the Lagrangian ``manifold" (curve) $\Lambda =\{p=S^{'}(x)\}$ of the WKB function. 
As $P$ approaches toward $\Lambda$, the chord $QR$ becomes
to the tangent of $\Lambda$ and $\si_{0}(x,p)\rightarrow 0$ .
It is clear that the two stationary  points of (\ref{alpha}) coalesce to the double point $\si=0$ as $(x,p)$ moves towards $\Lambda$.

Since the ordinary stationary-phase formula (\ref{statphform}) fails for the integral (\ref{wigwkb})
since\\ $\partial_{\si\si}F(\si=0,x,p)=0$,  we must to apply the uniform stationary formula  (\ref{ap4}). For applying this formula, we need first  to identify the small parameter $\alpha$, which controls the distance between the stationary points $\pm\si_{0}(x,p)$ of the Wigner phase. In order to do this, we expand $F$ in Taylor series about $\si=0$ ,
\begin{eqnarray*}
F(\si,x,p)&=& S(x)+\si S^{'}(x)+\frac{\si^2}{2}S^{''}(x)+\frac{\si^3}{6}
S^{'''}(x)+\dots\\
&&- \left(S(x)-\si
S^{'}(x)+\frac{\si^2}{2}S^{''}(x)-\frac{\si^3}{6}S^{'''}(x)+\dots \right)-2p\si\\
&=&-2(p-S^{'}(x))\si+\frac{1}{3}S^{'''}(x)\si^3+O(\si^5) \ .
\end{eqnarray*}
It becomes evident that for $P$ lying close enough to $\Lambda$,
the parameter $\alpha$ has to be identified as
\begin{equation}\label{parameter_alpha}
\alpha=\alpha(x,p):=p-S^{'}(x) \ ,
\end{equation}
since by
\begin{eqnarray*}
\partial_{\si\si}F(\si,x,p)= -2(p-S^{'}(x)) + S^{'''}(x)\si^2+O(\si^4) \ ,
\end{eqnarray*}
we easily see that $\si=0$ is a double stationary point for $p=S^{'}(x)$ .
Then, for any fixed $x$, we rewrite the Wigner phase $F$ in the
form\begin{eqnarray}
F(\si,\alpha,x)&=& S(x+\si)-S(x-\si)-2\si(\alpha+S^{'}(x))\nonumber\\
&=& \left (S(x+\si)-S(x-\si)-2\si S
^{'}(x)\right )-2\si\alpha \quad ,
\end{eqnarray}
and we have
\begin{equation}
\partial_{\si\si}F(\si=0,x,p)=0\quad , \quad  \partial_{\si\si\si}F(\si=0,x,p)=2S^{'''}(x)\quad , \quad
F_{\si\alpha}(\si=0,\alpha,x)=-2 \neq 0\ .
\end{equation}
These are exactly the conditions  on the phase which are required
for applying the uniform asymptotic formula (\ref{ap4}), and we get the approximation
\begin{eqnarray}\label{sclwigairy}
{\mathcal{W}}^{\epsilon}(x,p)\approx {\widetilde{\mathcal{W}}}^\epsilon(x,p)&:=& \frac{2^{2/3}}{\epsilon^{2/3}}\left(\frac{2}{\mid
S^{'''}(x)\mid} \right)^{1/3} A^{2}(x) \nonumber \\
&&\times Ai\left(-\frac{2^{2/3}}{\epsilon^{2/3}}\left(\frac{2}{S^{'''}(x)} \right)^{1/3}(p-S'(x))\right)     \ ,  \nonumber\\
\end{eqnarray}
which holds simultaneously for small $\epsilon$,  and  $(x,p)$ near the Lagrangian curve $\Lambda =\{p=S^{'}(x)\}$  of the WKB function.

We refer to the phase-space function ${\widetilde{\mathcal{W}}}^\epsilon(x,p)$ as  {\emph{the semiclassical Wigner function}}, corresponding to the WKB function (\ref{wkb_1}).


\bibliographystyle{amsplain}

\end{document}